\documentclass[a4paper,fleqn,article,usenatbib]{mnras} 
 
\usepackage[T1]{fontenc} 
\usepackage{ae,aecompl} 
\usepackage{array} 
\usepackage{graphicx}
\newcolumntype{C}{>{\centering\arraybackslash}m{1.9cm}}

\title[Supernovae 2009ip-like]{Supernovae 2016bdu and 2005gl, and their link with SN~2009ip-like transients: another piece of the puzzle} 
 
\author[A. Pastorello et al.]{ 
A. Pastorello,$^{1}$\thanks{E-mail: andrea.pastorello@oapd.inaf.it} 
C. S. Kochanek,$^{2,3}$ 
M. Fraser,$^{4,5,6}$ 
S. Dong,$^{7}$ 
N. Elias-Rosa,$^{1}$ 
\newauthor  
S. Benetti,$^{1}$ 
E. Cappellaro,$^{1}$ 
L. Tomasella,$^{1}$ 
A. J. Drake,$^{8}$ 
J. Hermanen,$^{9}$ 
\newauthor 
T. Reynolds,$^{9,10}$ 
B. J. Shappee,$^{11,12}$ 
S. J. Smartt,$^{13}$ 
K.~C. Chambers,$^{14}$ 
\newauthor 
M. E. Huber,$^{14}$ 
K. Smith,$^{13}$ 
K.~Z. Stanek,$^{2,3}$ 
A.~V. Filippenko,$^{15}$ 
E.~J. Christensen,$^{16}$ 
\newauthor 
L. Denneau,$^{14}$ 
S. G. Djorgovski,$^{8}$ 
H. Flewelling,$^{14}$ 
C. Gall,$^{17,18}$ 
A. Gal-Yam,$^{19}$ 
\newauthor 
S. Geier,$^{20,21}$ 
A. Heinze,$^{14}$ 
T.~W.-S. Holoien,$^{2,3,22}$ 
J. Isern,$^{23}$ 
T. Kangas,$^{9}$ 
\newauthor  
E. Kankare,$^{13}$ 
R.~A. Koff,$^{24}$ 
J.-M. Llapasset,$^{25}$ 
T.~B. Lowe,$^{14}$ 
P. Lundqvist,$^{26}$ 
\newauthor  
E. A. Magnier,$^{14}$ 
S. Mattila,$^{9}$ 
A. Morales-Garoffolo,$^{27}$ 
R. Mutel,$^{28}$ 
J. Nicolas,$^{29}$ 
\newauthor  
P. Ochner,$^{1,30}$ 
E. O. Ofek,$^{19}$ 
E. Prosperi,$^{31}$ 
A. Rest,$^{32}$ 
Y. Sano,$^{33,34,35}$ 
B. Stalder,$^{14}$ 
\newauthor 
M. D. Stritzinger,$^{18}$ 
F. Taddia,$^{26}$ 
G. Terreran,$^{1,13,30}$ 
J. L. Tonry,$^{14}$ 
R. J. Wainscoat,$^{14}$ 
\newauthor 
C. Waters,$^{14}$ 
H. Weiland,$^{14}$ 
M. Willman,$^{14}$ 
D.~R. Young,$^{13}$ 
and W. Zheng$^{15}$\\ 
\\ 
$^{1}$INAF-Osservatorio Astronomico di Padova, Vicolo dell'Osservatorio 5, I-35122 Padova, Italy \\ 
$^{2}$Department of Astronomy, The Ohio State University, 140 West 18th Avenue, Columbus, OH 43210, USA\\ 
$^{3}$Center for Cosmology and AstroParticle Physics (CCAPP), The Ohio State University, 191 W. Woodruff Ave., Columbus, OH 43210, USA \\ 
$^{4}$School of Physics, O'Brien Centre for Science North, University College Dublin, Belfield, Dublin 4, Ireland\\ 
$^{5}$Institute of Astronomy, University of Cambridge, Madingley Road, CB3 0HA, Cambridge, UK\\ 
$^{6}$Royal Society -- Science Foundation Ireland University Research Fellow\\ 
$^{7}$Kavli Institute for Astronomy and Astrophysics, Peking University, Yi He Yuan Road 5, Hai Dian District, Beijing 100871, China\\ 
$^{8}$Astronomy Department, California Institute of Technology, Pasadena, CA 91125, USA \\ 
$^{9}$Tuorla Observatory, Department of Physics and Astronomy, University of Turku, V\"ais\"al\"antie 20, FI-21500 Piikki\"o, Finland\\ 
$^{10}$Nordic Optical Telescope, Apartado 474, E-38700 Santa Cruz de La Palma, Spain\\ 
$^{11}$Carnegie Observatories, 813 Santa Barbara Street, Pasadena, CA  91101, USA\\ 
$^{12}$Hubble Fellow and Carnegie-Princeton Fellow\\ 
$^{13}$Astrophysics Research Centre, School of Mathematics and Physics, Queen's University Belfast, Belfast BT7 1NN, UK\\ 
$^{14}$Institute for Astronomy, University of Hawaii, 2680 Woodlawn Drive, Honolulu, HI 96822, USA\\ 
$^{15}$Department of Astronomy, University of California, Berkeley, CA 94720-3411, USA\\ 
$^{16}$Lunar and Planetary Lab, Department of Planetary Sciences, University of Arizona, Tucson, AZ 85721, USA\\ 
$^{17}$Dark Cosmology Centre, Niels Bohr Institute, University of Copenhagen, Juliane Maries Vej 30, 2100 Copenhagen, Denmark\\ 
$^{18}$Department of Physics and Astronomy, Aarhus University, Ny Munkegade 120, DK-8000 Aarhus C, Denmark\\ 
$^{19}$Department of Particle Physics and Astrophysics, Faculty of Physics, The Weizmann Institute of Science, Rehovot 76100, Israel\\ 
$^{20}$Gran Telescopio Canarias (GRANTECAN), Cuesta de San Jos\'e s/n, E-38712, Bre\~na Baja, La Palma, Spain\\ 
$^{21}$Instituto de Astrof\'isica de Canarias, V\'ia L\'actea s/n, E-38200, La Laguna, Tenerife, Spain\\ 
$^{22}$US Department of Energy Computational Science Graduate Fellow\\ 
$^{23}$Institut de Ci\`encies de l'Espai (CSIC - IEEC), Campus UAB, Cam\`i de Can Magrans S/N, E-08193 Cerdanyola (Barcelona), Spain \\ 
$^{24}$Antelope Hills Observatory, 980 Antelope Drive West, Bennett, CO 80102, USA\\ 
$^{25}$66 Cours de Lassus, F-66000 Perpignan, France\\ 
$^{26}$Department of Astronomy and The Oskar Klein Centre, AlbaNova University Center, Stockholm University, SE-10691 Stockholm, Sweden\\ 
$^{27}$Applied Physics Department, Polytechnic Engineering School of Algeciras, University of C\'adiz, Avenida Ram\'on Puyol S/N, 11202 \\ Algeciras, Spain \\ 
$^{28}$Department of Physics and Astronomy, University of Iowa, Iowa City, IA 52242, USA\\ 
$^{29}$Groupe SNAude France, 364 Chemin de Notre Dame, 06220 Vallauris, France\\ 
$^{30}$Dipartimento di Fisica e Astronomia, Universit\`a di Padova, via Marzolo 8, I-35131 Padova, Italy\\ 
$^{31}$Osservatorio Astronomico di Castelmartini, Via Bartolini 1317, 51036 Larciano, Pistoia, Italy\\ 
$^{32}$Space Telescope Science Institute, 3700 San Martin Drive, Baltimore, MD 21218, USA\\ 
$^{33}$Observation and Data Center for Cosmosciences, Faculty of Science, Hokkaido University, Kita-ku, Sapporo, Hokkaido 060-0810, Japan\\ 
$^{34}$Nayoro Observatory, 157-1 Nisshin, Nayoro, Hokkaido 096-0066, Japan\\ 
$^{35}$VSOLJ, Nishi juni-jou minami 3-1-5, Nayoro, Hokkaido 096-0022, Japan\\ 
} 
 
\date{Accepted XXX. Received YYY; in original form ZZZ} 
 
\pubyear{201X} 
 
\begin{document} 
\label{firstpage} 
\pagerange{\pageref{firstpage}--\pageref{lastpage}} 
\maketitle 
 
\clearpage 
 
\begin{abstract} 
 
Supernova (SN) 2016bdu is an unusual transient resembling SN~2009ip. SN~2009ip-like events are characterized by a long-lasting phase of  
erratic variability which ends with two luminous outbursts a few weeks apart. The second outburst is significantly more luminous (about 3 mag)  
than the first. In the case of SN~2016bdu, the first outburst (Event A) reached an absolute magnitude $M_r \approx -15.3$ mag, while the second  
one (Event B) occurred over one month later and reached $M_r \approx$ $-18$ mag. By inspecting archival data, a faint source at the position  
of SN~2016bdu is detectable several times in the past few years. We interpret these detections as signatures of a phase of erratic variability,  
similar to that experienced by SN~2009ip between 2008 and mid-2012, and resembling the currently observed variability of the luminous blue variable  
SN~2000ch in NGC~3432. Spectroscopic monitoring of SN~2016bdu during the second peak initially shows features typical of a SN~IIn.  
One month after the Event B maximum, the spectra develop broad Balmer lines with P~Cygni profiles and broad metal features. At these  
late phases, the spectra resemble those of a typical Type II SN.  All members of  this SN~2009ip-like group are remarkably similar to the Type  
IIn SN~2005gl. For this object, the claim of a terminal SN explosion is supported by the disappearance of the progenitor star. The similarity with  
SN~2005gl suggests that all members of this family may finally explode as genuine SNe, although the unequivocal detection of  
nucleosynthesised elements in their nebular spectra is still missing. 
 
\end{abstract} 
 

 
\section{Introduction} 
 
While it is well known that many massive stars lose a large fraction of their envelope in the latest stages of their life,  
the mechanisms that trigger the mass loss are still poorly understood. Steady winds \citep{cas75,owo99,dwa02,lam02,che11,mor11,gin12},  
enhanced mass loss due to binary interaction \citep{kas10a,kas10,smi11,che12,sok12,sok13,sel13}, or major outbursts caused by stellar  
instabilities \citep{hum94,lan99,woo07,arn11,chat12,mor14,shi14} can all lead to the formation of extended circumstellar environments 
(for reviews on this topic, see \citealt{lan12} and \citealt{smi14a}).
When stars embedded in such dense cocoons explode as supernovae (SNe), they produce  
the typical observables of interacting SNe: narrow to intermediate-width lines in emission, a blue spectral continuum, and  enhanced X-ray,  
ultraviolet (UV), and radio fluxes \citep{wei86,chu91,che94,fil97,are99,kie12,cha12,cha15,smi17}.  
SNe showing spectra with narrow or intermediate-width Balmer lines produced in an H-rich circumstellar medium (CSM) are classified as Type IIn \citep{sch90}, while  
those whose spectra are dominated by narrow or intermediate-width He~I lines are classified as Type Ibn \citep{pasto08,pasto16,hos17}. 
 
Signatures of major instabilities in the last stages of life of very massive stars are frequently observed.  
These nonterminal eruptions are usually labelled as ``SN impostors'' \citep{van00,mau06}. Although they do not necessarily  
herald terminal SN explosions on short timescales, such eruptions have been detected from a few weeks to years  
before the SN in some cases. A seminal case is the Type Ibn SN 2006jc, which had a luminous outburst two years before 
the final explosion  \citep{pasto07,fol07}. Moderate-intensity pre-SN outbursts were also likely observed in more canonical  
stripped-envelope SNe \citep{cor14,str15}. 
Much more common is evidence of pre-SN bursts from Type IIn SN progenitors  
\citep[e.g.,][]{fra13a,ofek14}. Pre-SN outbursts were well observed in two Type IIn events: SN~2009ip  
\citep{mau13,mau14,pasto13,fra13b,ofe13a,fra15,mar14,gra14,smi13,smi16,gra17} and SN~2015bh \citep{ofek16,ner16,tho16,gor16}.  
Both sources had historical light curves with signatures of erratic variability over timescales of years (SN impostor phase),  
followed by two luminous outbursts separated by a few weeks. In each case, the first outburst (labelled as ``Event A'') had an absolute  
magnitude $M_R \approx$  $-15$ mag, and the second one (``Event B'') was brighter, approaching or exceeding $M_R \approx -18$ mag.   
From a careful inspection of the light curve of SN~2009ip after the Event B maximum, \citet{gra14} and \citet{mar15} 
noted luminosity fluctuations consistent with the ejecta colliding with CSM shells produced 
during the earlier eruptive phase.  
 
Light curves with two outbursts were also observed in SN~2010mc \citep{ofek13} and LSQ13zm \citep{leo16}. These transients were discovered  
in relatively distant galaxies, and their previous SN-impostor-phase variability would have been too faint to be observed by existing surveys. 
The Type IIn SNhunt151 in UGC~3165 was another object showing a double-outburst light curve, but  had also significant spectroscopic differences from  
typical SN~2009ip-like transients \citep{nancy17}.   
More recently, SN~2016cvk and SN~2016jbu were also  announced as possible SN~2009ip-like candidates \citep{bro16,fra17,bos17,kil17}. 
 
Even accounting for these recent discoveries, pre-SN outbursts have been directly detected only occasionally. Nonetheless, based on the  
Palomar Transient Factory (PTF) sample control time and coadding images in multiple-day bins to go deeper than the nominal limiting magnitude  
of the survey, \citet{ofek14} claim that these events are quite frequent, but in most  cases below the detection threshold of pre-explosion images.  
This likely explains the lack of outburst precursors in the sample of SNe~IIn of the Katzman Automatic Imaging Telescope survey \citep{bil15}. 
 
Some authors \citep[e.g.,][]{pasto13,tho16} have noted the resemblance of the impostor phase of SNe 2009ip and 2015bh  
to the erratic variability exhibited by the SN impostor NGC3432-LBV1 
\citep[also known as SN~2000ch;][]{wag04,pasto10} over the last two decades. This is in fact an excellent candidate to become another SN~2009ip analog,  
perhaps within a human lifetime. 
 
In this context, a new SN~2009ip-like event is important for improving our understanding of these unusual explosions.  
SN~2016bdu was discovered on 2016 May 24.43 UT (JD = 2,457,532.93; UT dates are used throughout this paper) by the PanSTARRS-1 (PS1) Survey for Transients (PSST)\footnote{Oddly,  
the temporary PSST name, PS16bdu, recalled the final IAU designation.} \citep{hub15,cha16} at  $\alpha = 13^h10^m13.95^s$,  
$\delta = +32^\circ31'14.07''$ (J2000.0).  
The SN candidate exploded in the very faint ($g = 21.19$, $r = 20.94$ mag) galaxy SDSS~J131014.04+323115.9. 
The SN is clearly offset by $2.1''$ from the center of its host  in the PS1 images. 
The All-Sky Automated Survey for Supernovae \citep[ASAS-SN;][]{sha14} detected the object again on 2016 May 29.38,  
and soon thereafter it was classified by the 2.56~m Nordic Optical telescope (NOT) Unbiased Transient Survey (NUTS)\footnote{{http://csp2.lco.cl/not/.}}  
collaboration \citep{mat16} as a Type IIn SN \citep{giac16a}.  
 
\begin{figure} 
	\includegraphics[width=\columnwidth]{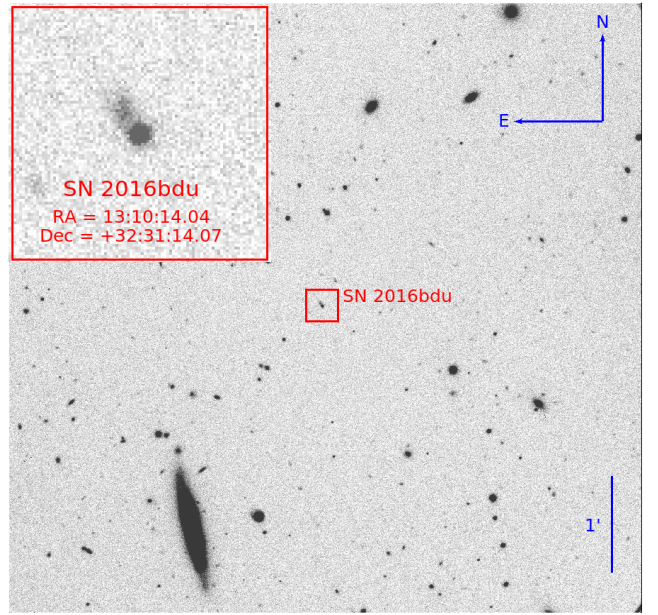} 
    \caption{SN~2016bdu and the surrounding stellar field. The large, edge-on spiral galaxy visible in the bottom-left corner is UGC 08250.  
Sloan $r$-band image taken on 2016 February 29 with the NOT and the ALFOSC camera. A detail showing the host galaxy and the SN is 
in the upper-left corner.} 
    \label{fig1} 
\end{figure}

 \citet{giac16a} noted that the transient was offset by $2.6'$ (hence about 57 kpc) from the center 
of a relatively large, edge-on spiral galaxy, UGC 08250. This galaxy has a redshift $z = 0.0176$ \citep{gar15,cou15}. 
For H$_0$ = 73 km s$^{-1}$ Mpc$^{-1}$, $\Omega_{M} = 0.27$, and $\Omega_{\Lambda} = 0.73$, the luminosity distance of $d_L = 73.3$ Mpc and 
the distance modulus of $\mu = 34.33$ mag. If we correct for Local Group infall into Virgo ($v_{\rm Vir} = 5474$ km s$^{-1}$), we infer 
a slightly larger luminosity distance of $d_{L, {\rm Vir}} = 76.1$ Mpc ($\mu_{\rm Vir} = 34.41$ mag).  
We will assume that the faint host of SN~2016bdu (SDSS~J131014.04+323115.9) 
is associated with UGC 08250 (see Figure \ref{fig1}). From the position of the peaks of the  
most prominent narrow emission lines in the transient's spectra, we estimate the redshift of SDSS~J131014.04+323115.9 to be  
$z = 0.0173 \pm 0.0002$ ($d_L = 72.0$ Mpc and $\mu = 34.29$ mag), consistent with this hypothesis. 
We adopt the Virgo-infall-corrected distance modulus for this redshift of $\mu_{\rm Vir} = 34.37 \pm 0.15$ mag. 
 
The nondetection of  the interstellar Na~I doublet (Na~I~D) absorption lines at the redshift of SDSS~J131014.04+323115.9 suggests that there 
is negligible reddening due to the host galaxy; hence, 
we adopt only the Milky Way contribution $E(B-V) = 0.013$ mag \citep{sch11} as the total interstellar extinction toward SN~2016bdu. 
Given these distance and reddening estimates, SDSS~J131014.04+323115.9 has a total absolute magnitude $M_g = -13.23$ mag, and 
an intrinsic colour of $g-r = 0.23$ mag. This makes the galaxy hosting SN~2016bdu much less luminous than the Magellanic Clouds, possibly suggesting 
that SN~2016bdu exploded in a low-metallicity environment. 
 
The structure of this paper is as follows. In Sect. \ref{obs} we present photometric and spectroscopic observations of SN~2016bdu, 
Sect. \ref{discussion} discusses plausible scenarios to explain the sequence of outbursts experienced by the SN~2016bdu progenitor, and a summary follows  
in Sect. \ref{concl}. 
 
\section{Observations} \label{obs} 
 
Soon after the discovery, our inspection of archival images revealed that the prediscovery photometric evolution 
of the stellar precursor of SN~2016bdu resembled that of SN~2009ip. For this reason, we decided to initiate an extensive  
follow-up campaign in the optical and near-infrared (NIR) domains.

The optical and NIR data were obtained with the NOT using ALFOSC and NOTCam,  the 2.0~m Liverpool Telescope (LT) using IO:O, the 10.4~m  
Gran Telescopio Canarias (GTC) using OSIRIS, the 1.82~m Copernico Telescope using AFOSC, and the 1.5~m Tillinghast Telescope using the FAST spectrograph.  
Additional photometry was obtained using a Meade 10\arcsec ~LX-200 Schmidt-Cassegrain Telescope with an Apogee AP-47 CCD camera located near Bennett (Colorado, USA), 
and the 0.51~m Iowa Robotic Telescope of the Winer Observatory (in southern Arizona, USA), equipped with a  
cooled, back-illuminated 1024 $\times$ 1024 pixel CCD sensor. 
Further photometry (including archival data) was later provided by  
ASAS-SN\footnote{The survey uses four 14~cm ``Brutus'' robotic telescopes located  
in the Haleakala station (Hawaii, USA) of the Las Cumbres Observatory Global Telescope (LCOGT) network.} \citep{sha14}; 
PSST\footnote{PSST uses the PS1-1.8~m telescope \protect\citep{cha16}, which has a 7 square degree field of view, with a mosaic 
CCD camera, operating on Haleakala in the island of Maui, Hawaii, USA.} \citep{hub15}; the Catalina Real-Time Transient Survey\footnote{The 
survey uses the 0.7~m Schmidt telescope of the Bigelow Station, and has an archive covering about 13~yr of observations.} \citep[CRTS,][]{dra09,djo12}; and 
the Asteroid Terrestrial-impact Last Alert System (ATLAS)\footnote{This survey uses two 0.5~m wide-field telescopes on Mauna Loa and Haleakala 
in Hawaii, USA \protect\citep{ton11}, one of which is operational.}. Additional $R$-band data were provided by the  
PTF second data release \citep{law09,ofek12}, from the Infrared Processing and Analysis  
Center\footnote{http://www.ipac.caltech.edu/; the PTF survey used the 1.2~m Oschin Telescope at Palomar Observatory  
equipped with a 7.8 square degree CCD array (CFH12K).} \citep{lah14}. Finally, a few photometric epochs from 1998 and 2003  
calibrated in the Johnson-Bessell $V$-band magnitude scale were obtained from images taken by the Near Earth Asteroid Tracking  
(NEAT) program, and retrieved through the  SkyMorph GSFC website.\footnote{http://skyview.gsfc.nasa.gov/skymorph/skymorph.html.} 
 
\subsection{Photometry} \label{photo} 
 
\begin{figure} 
	\includegraphics[width=\columnwidth]{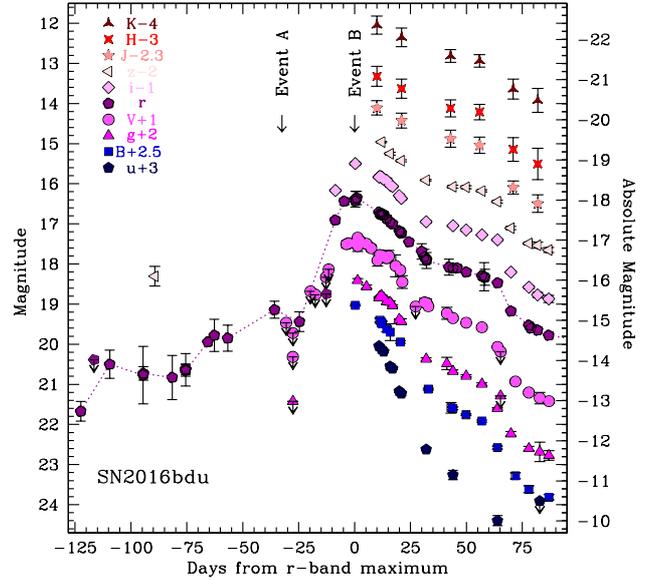} 
    \caption{Multi-band light curves of SN~2016bdu. The most recent prediscovery observations are also shown. The phases are  
computed with respect to the Sloan $r$-band maximum of Event B (JD = 2,457,541.5; see text). The epochs of the peaks of Events A and B 
are indicated. The $BVJHK$ data are in the Vega magnitude system, and the Sloan $ugriz$ data are in the AB magnitude system. The dotted line connects 
$r$-band data, and reveals fluctuations in the light curve during Event A and in the post-peak decline from Event B.} 
    \label{fig2} 
\end{figure} 
 
The science images were first reduced using IRAF\footnote{IRAF is distributed by the National Optical Astronomy Observatory, which is operated  
by the Association of Universities for Research in Astronomy (AURA) under a cooperative agreement with the National Science Foundation (NSF). }. These  
preliminary operations included overscan, bias, and flat-field corrections, for both imaging and spectroscopy. The magnitudes of SN~2016bdu were  
measured using a dedicated package \citep[SNOoPY;][]{cap14} that performs point-spread-function (PSF) fitting photometry on the original or the  
template-subtracted images. Since the SN field falls in the sky area mapped by the Sloan Digital Sky Survey (SDSS), we identified a sequence of reference  
stars, and measured nightly zero-points and instrumental colour terms. These were used to accurately calibrate the SN magnitudes on the different  
nights. The Johnson-Bessell $B$ and $V$ magnitudes of the reference stars were computed from the Sloan magnitudes following the relations of \citet{cho08}.  
PTF $R$-band data were converted to the Sloan $r$-band photometric system using magnitudes of comparison stars taken from the SDSS catalogue,  
while unfiltered data were scaled to Sloan $r$-band magnitudes.  
NIR images from NOTCam were reduced using a slightly modified version of the IRAF  
package NOTCam v.2.5\footnote{http://www.not.iac.es/instruments/notcam/guide/observe.html\#reductions.}, and photometric measurements were  
performed after the subtraction of the luminous NIR sky. The instrumental SN magnitudes were calibrated using the 2MASS catalogue \citep{skr06}. 
Final SN magnitudes are listed in Tables \ref{tab:2016bdu_JB}, \ref{tab:2016bdu_SDSS}, and \ref{tab:2016bdu_NIR} of Appendix \ref{app1a}.

\begin{figure*} 
	\includegraphics[width=17.8cm]{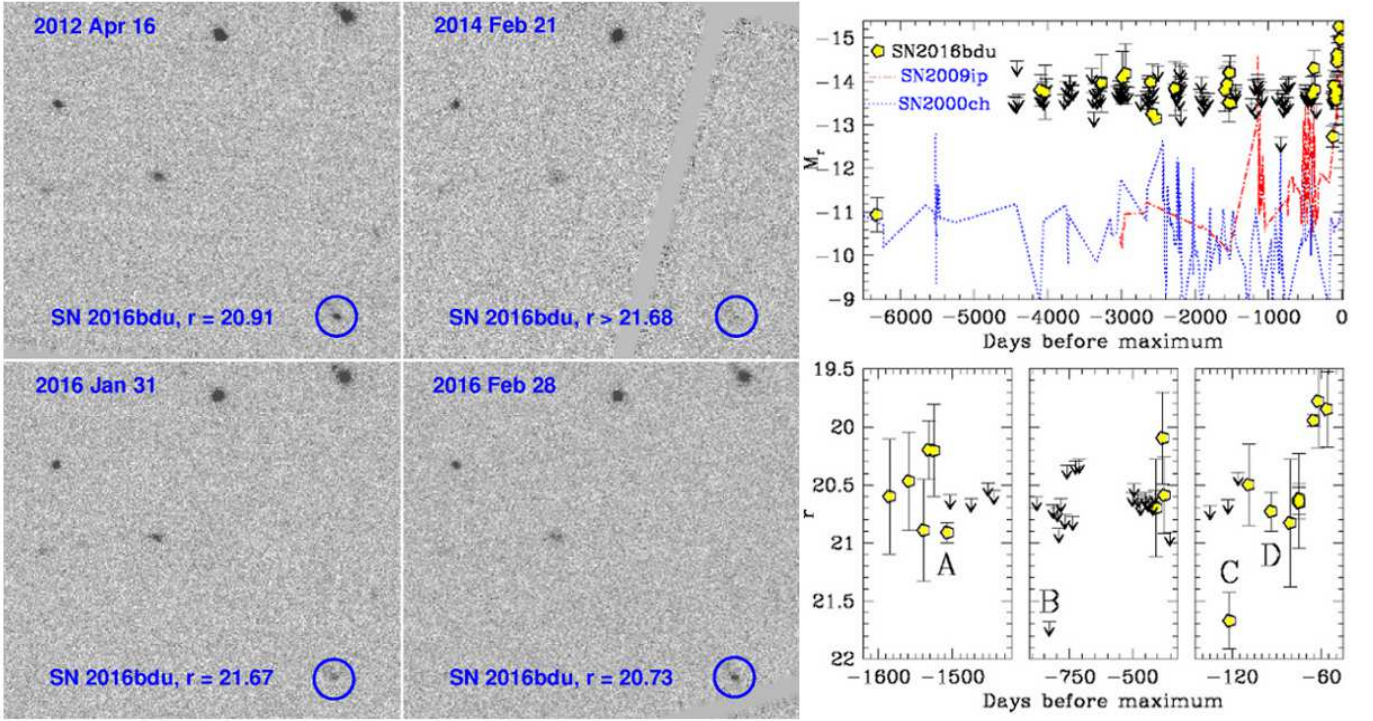} 
    \caption{ Left panel: Prediscovery Sloan $r$-band images of the location of SN~2016bdu obtained with PS1 on 2012 April 16 (top left, A), 2014 February 21 (top right, B),  
2016 January 31 (bottom left, C), 2016 February 28 (bottom right, D), with the transient being at different luminosities. Right panel, top: comparison of the prediscovery absolute $r$-band  
light curve of SN~2016bdu with the $R$-band light curves of the impostor SN 2000ch and the pre-explosion variability of SN~2009ip. Right panel, bottom: $r$-band magnitude 
variability of the transient in SDSS~J131014.04+323115.9 before its discovery. The magnitudes corresponding to the images on the left are marked with uppercase letters.} 
    \label{fig3} 
\end{figure*}

\begin{figure*} 
	\includegraphics[width=18cm]{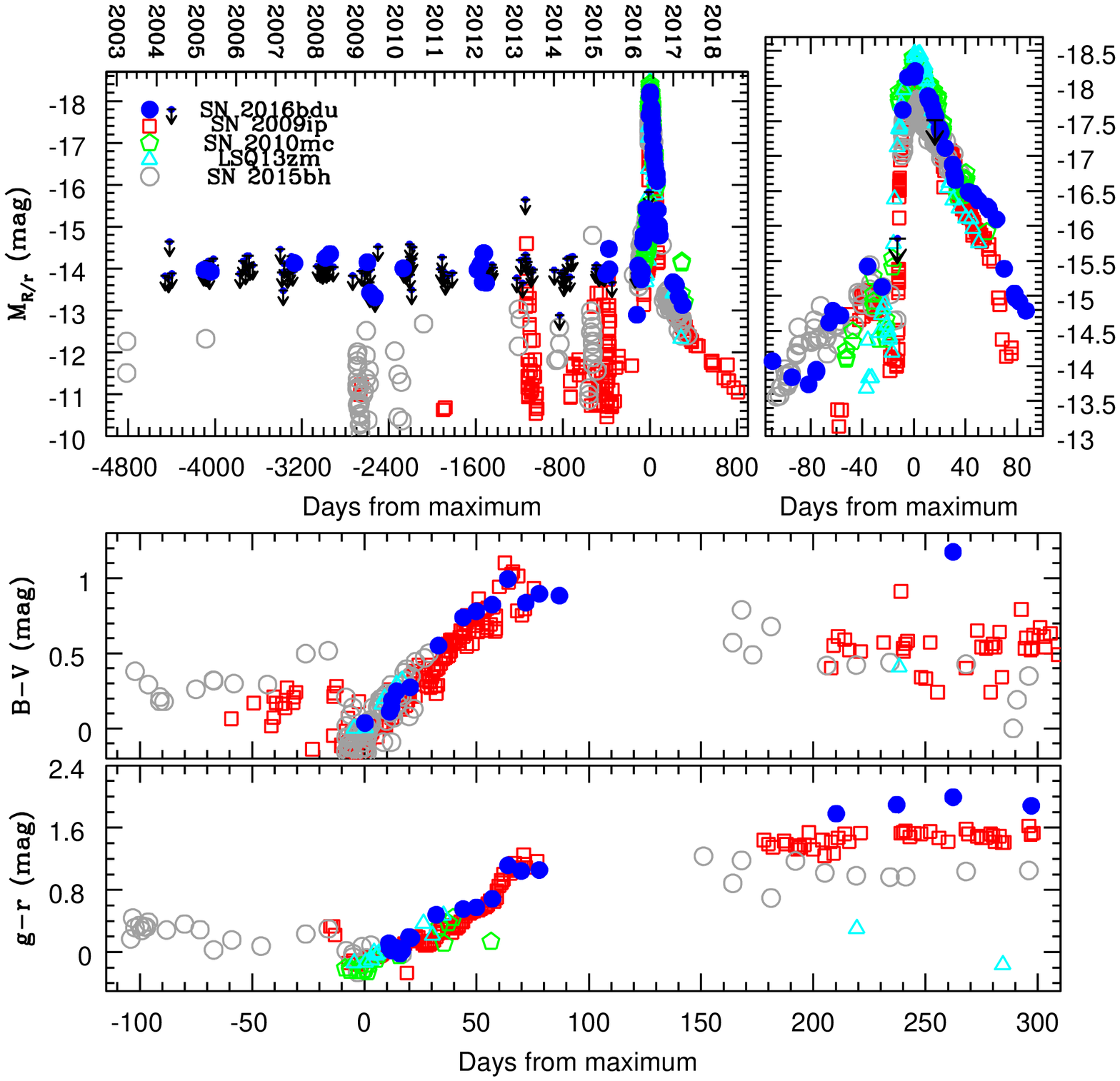} 
    \caption{Johnson-Cousins $R$ or Sloan $r$ absolute magnitude light curves for our sample of SN~2009ip-like events, spanning 15~yr (top left),  
and during the period encompassing the Events A and B (top right).  
For clarity, prediscovery detection limits are shown only for SN~2016bdu. These limits were obtained by placing an artificial star close 
to the position of SN~2016bdu with a typical signal-to-noise ratio (S/N) of 1.5--2.5. 
The years above the top-left panel refer to the evolution of SN~2016bdu. 
$B-V$ (middle) and $g-r$ (bottom) colour evolution for SN~2009ip-like SNe. The phases are in days from the Event B peak.} 
    \label{fig4} 
\end{figure*} 
 
As shown in Figure \ref{fig2}, the light curve of SN 2016bdu has two main brightening events, similar to other 
SN~2009ip-like transients. In analogy with the labelling adopted for SN~2009ip \citep{pasto13}, the nonmonotonic brightening observed  
in the light curve from about four months before the discovery of SN~2016bdu is named Event A. We note that Event A of SN~2009ip is different, 
having a much shorter duration and a monotonic rise to the maximum. In SN~2016bdu, Event A reaches a peak of $r = 19.1 \pm 0.2$ mag  
(on JD = 2,457,509 $\pm 6$, obtained through a low-order polynomial fit) in about three months. Then the luminosity slightly declines to  
$r = 19.44 \pm 0.25$ mag, before rising again to the second peak (Event B) which is reached about one month later (on JD = 2,457,541.5 $\pm 1.5$).  
The maximum magnitude of Event B is $r = 16.37 \pm 0.03$ mag. The $V$-band peak of Event B is reached on $JD = 2,457,540.9 \pm 1.8$, with 
$V = 16.46 \pm 0.03$ mag. After maximum, the light curve declines rapidly for about one month, more slowly between days 30 and 60 past-peak, 
and finally the decline rate increases again during the last month covered by our photometric campaign. This trend is observed both in the  
optical and NIR bands. As mentioned before, low-contrast undulations are observed in all bands during Event A (see Figure \ref{fig2}),  
and when the light curve declines after the peak of Event B. Similar behaviour was observed in SN~2009ip \citep{gra14,mar15} and interpreted  
as signatures of ejecta colliding with previously ejected circumstellar shells. 
 
In this context, it is worth noting that some sparse detections of a source at the position of  SN~2016bdu have occurred over a  
period of several years before the SN (see Table \ref{tab:2016bdu_SDSS}). A few marginal detections in CRTS images are registered from 2005  
to 2008. We note that they are 0.3--0.5 mag brighter than the host-galaxy magnitude ($r = 20.94$ mag; this is fainter than the typical  
detection limits of individual CRTS images). A source is also visible near the SN position in April--June 2009, although the magnitudes in the  
PTF images ($r \approx  21.1$--21.3 mag) are consistent (within the uncertainties) with that of the host galaxy. From about four years before the SN  
discovery, there is a set of clear detections, with the source well resolved in the host galaxy in several PS1 images: in January--April 2012  
($r = 20.20 \pm 0.25$ mag), March 2014 ($g = 21.63 \pm 0.17$, $i = 21.44 \pm 0.29$ mag), and April--May 2015 ($r = 20.1 \pm 0.4$~mag).  
Finally, there are repeated detections, starting in January 2016, before the rise to the Event A peak. The prediscovery data are shown in  
Figure \ref{fig3}. Unfortunately, in most images collected from CRTS, the source is below the instrumental detection threshold. However, the object  
is detected in the deeper PS1 and PTF images, indicating that the object was in outburst (with absolute magnitude $M_R$ ranging from $-13$ to  
$-14$~mag), and was characterised by erratic variability. The peak magnitudes of these prediscovery images of SN~2016bdu are comparable with  
those of the brightest outbursts of SN~2009ip during the impostor phase \citep[years 2010--2012;][see also the top-right panel in Figure \ref{fig3}]{pasto13,mau13}. 
However, because of the larger distance of SN~2016bdu and the limited depth of most of the archival images, we generally have only upper limits 
on the source flux in the low-luminosity states between the outbursts. 
 
\begin{table*} 
	\centering 
	\caption{Spectroscopic observations of SN~2016bdu. Phases calculated from Sloan $r$-band max of Event B (JD = 2,457,541.5).} 
	\label{tab:spectra} 
	\begin{tabular}{cccccc}  
		\hline 
		Date & JD & Phase (d) & Instrumental config. & Range (\AA) & FWHM (\AA) \\ 
		\hline 
2016-06-02 & 2457542.49 & 1.0 & NOT + ALFOSC + gm4 & 3500--9700 & 14 \\ 
2016-06-03 & 2457542.68 & 1.2 & Tillinghast + FAST + 300tr & 3500--7400 & 6 \\ 
2016-06-14 & 2457553.55 & 12.1 & NOT + ALFOSC + gm4 & 3500--9700 & 18 \\   
2016-06-17 & 2457557.37 & 15.9 & 1.82~m Copernico + AFOSC + VPH7 + VPH6 & 3350--9300 & 14 \& 15 \\ 
2016-06-22 & 2457562.39 & 20.9 & 1.82~m Copernico + AFOSC + VPH7 + VPH6 & 3500--9300 & 14 \& 15 \\ 
2016-07-02 & 2457571.51 & 30.0 & NOT + ALFOSC + gm4 & 3400--9700 & 14 \\ 
2016-07-15 & 2457585.45 & 44.0 & NOT + ALFOSC + gm4 & 3400--9650 & 18 \\ 
2016-07-17 & 2457587.43 & 45.9 & GTC + OSIRIS + R1000B + R1000R & 3650--9350 & 7 \& 8 \\ 
2016-07-29 & 2457599.46 & 58.0 & NOT + ALFOSC + gm4 & 3400--9700 & 14 \\ 
2016-08-12 & 2457613.41 & 71.9 & TNG + LRS + LRB & 3500--8000 & 14 \\ 
2016-08-19 & 2457620.38 & 78.9 & GTC + OSIRIS + R1000R & 5100--9600 & 8 \\ 
2017-01-17 & 2457770.73 & 229.2 & GTC + OSIRIS + R1000R & 5100--9300 & 8 \\ 
2017-01-20 & 2457773.68 & 232.2 & GTC + OSIRIS + R1000B + R1000R & 3700--10,100 & 7 \& 8 \\ 
        	\hline 
        \end{tabular} 
\end{table*} 
 
In Figure \ref{fig4}, we compare the $r$-band absolute magnitude light curve of SN~2016bdu to those of SN~2009ip \citep{smi10,pasto13,mau13,pri13,fra13b,fra15,mar14,gra14,mar15}, SN~2010mc  
\citep{ofek13}, LSQ13zm \citep{leo16}, and SN~2015bh \citep{ofek16,ner16,gor16,tho16}. If Sloan-$r$ data are not available, 
the Johnson-Cousins $R$-band light curves are shown. For homogeneity, all data have been transformed to the Vega system.  
The data for SN~2009ip, SN~2010mc, and LSQ13zm are corrected for Milky Way reddening. For SN~2015bh, we adopt the total reddening estimate of $E(B-V) = 0.21$ mag from \citet{tho16}. 
In the top-left panel, in particular, we show the long-term photometric evolution of our SN sample over a temporal window of almost 15~yr.  
A close-up view including the rise to Event A and the decline after the peak of Event B is shown in the top-right panel. 
The photometric evolution is surprisingly similar for all objects in this sample. In particular, we note the following. 
 
\begin{itemize} 
\item There are occasional detections in the ``impostor phase'' from weeks to years before the onset of Event A in at least three objects:  
SN~2009ip, SN~2015bh, and SN~2016bdu. For the two other objects, we speculate that the impostor phase remained undetected because of the larger  
distances to their host galaxies. 
\item The duration of Event A appears to vary for the five objects, lasting from a few weeks in LSQ13zm \citep{leo16} to over 3 months,  
 and with a bumpy rise in SN~2016bdu and SN~2015bh \citep{ner16,tho16}. The absolute magnitudes of the A events are between $M_{R/r} = -14.5$ and $-15.3$ mag. 
\item All objects show a relatively fast rise to the Event B maximum ($M_{r/R} \leq -18$ mag) followed by a relatively steep decline.  
Some undulations in the light curve of SN~2009ip are observed \citep{mar15}, and a nonlinear decline is also detected in SN~2016bdu. 
\item When late-time observations are available, the light curves have flattened, with decline rates slower than the $^{56}$Co decay  
and without any clear evidence of further light-curve fluctuations \citep{mar14,fra15,ner16,tho16}. 
\end{itemize} 
 
The bottom panel of Figure \ref{fig4} shows the $B-V$ and $g-r$  colours. Again, the evolution of the two colours is similar for all the  
objects from day $-20$ (from the peak of Event B) to day +80. In particular, the colours become bluer from day $-20$ to day 0, with $B-V$  
ranging from $-0.5$ to 0 mag, and $g-r$ from 0.3 to $-0.2$ mag. Later on, from day 0 to 80, the colours become redder, spanning from  
$B-V = 0$ to 1.1 mag and $g-r = -0.2$ to 1.2 mag. More dispersion in the colour evolution is  observed at later phases, especially in $g-r$,  
although  large uncertainties affect the photometric measurements at these epochs. Nonetheless, the $g-r$ colour of SN~2016bdu is the reddest  
in the sample, being about 2~mag at $\sim 8$ months after maximum brightness. 
 
\subsection{Quasibolometric light curve} \label{bolo} 
 
\begin{figure} 
	\includegraphics[width=\linewidth]{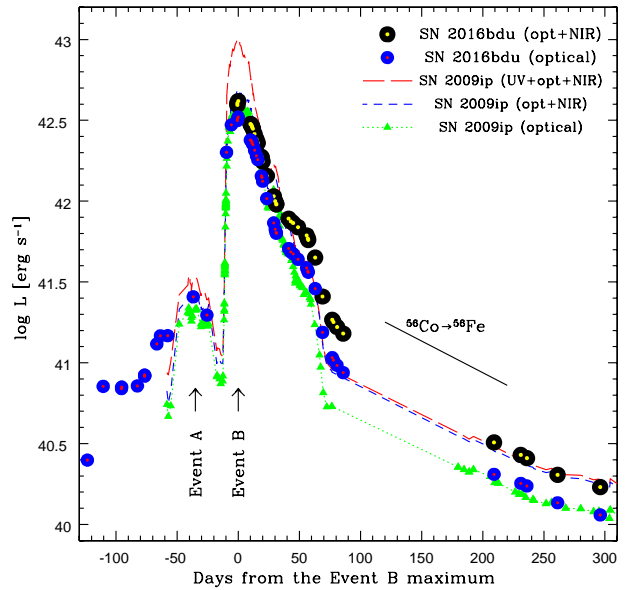} 
    \caption{Quasibolometric light curves of SNe~2016bdu and 2009ip. For SN~2016bdu, the curves obtained by integrating the optical bands only are shown as small blue-red points, and 
those including the NIR contribution are shown as large black-yellow points. For SN~2009ip, we show the optical curve (green dotted line), the optical plus NIR 
curve (blue, short-dashed line), and the {\it uvoir} curve (see text; red long-dashed line).} 
    \label{fig5} 
\end{figure} 
 
Using the available optical photometry of SN~2016bdu, we obtain a quasibolometric light curve by integrating the extinction-corrected fluxes at each epoch  
with the trapezoidal rule. We assume zero flux at the integration extremes. We also estimate a quasibolometric light curve including the NIR data when they  
are available. Since no NIR data are available for SN~2016bdu before the Event B peak, we compute the optical+NIR light curve only for later phases. The  
resulting quasibolometric light curves are shown in Figure \ref{fig5}. These curves are compared to those obtained for the best-studied example, SN~2009ip.  
For SN~2009ip, we show results for the optical bands only, optical plus NIR, and the {\it uvoir} (from the UV to the NIR domain). The similarity of the two  
objects is striking. In addition, the available data for SN~2016bdu suggest a late-time decline of its quasibolometric luminosity that is slightly flatter  
than the rate expected from the radioactive decay of $^{56}$Co into $^{56}$Fe. This suggests that the ejecta-CSM interaction is still the dominant mechanism  
powering the light curve of both SNe at very late phases. This claim will be discussed further in Sect. \ref{spec}. 
 
For SN~2016bdu, the NIR  contributes up to 20\% of the total luminosity budget at the Event B maximum, rising to about 30\% at the late phases ($\sim 8$  
months later). The missing UV contribution in SN~2016bdu can be estimated by assuming that it is similar to that of SN~2009ip, where the UV contribution is  
large (nearly 50\%) at the time of the Event B peak, and negligible ($<5$\%) at the late phases. 
 
\subsection{Spectroscopy} \label{spec} 
 
\begin{figure} 
	{\includegraphics[width=\linewidth]{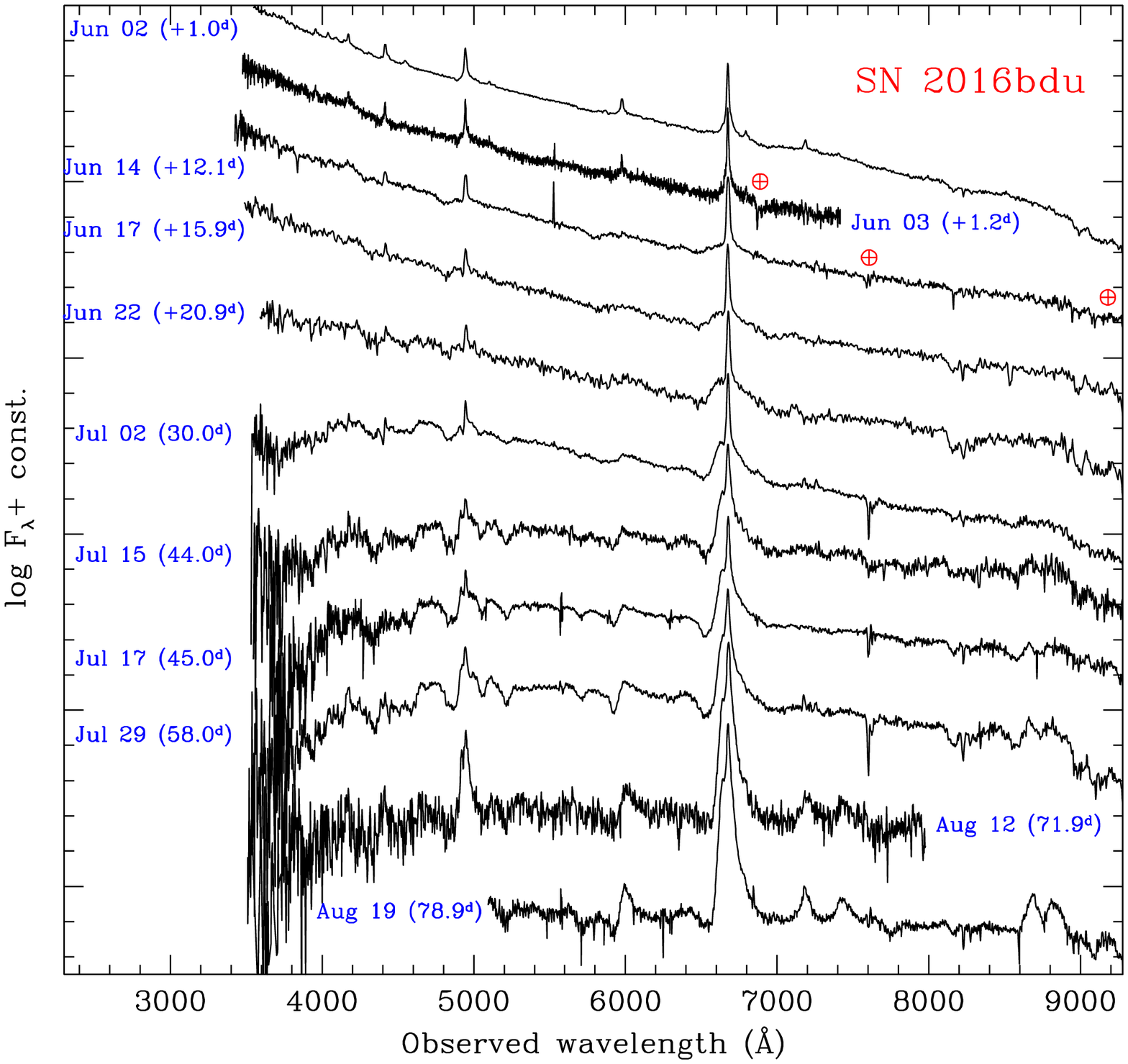} 
	\includegraphics[width=\linewidth]{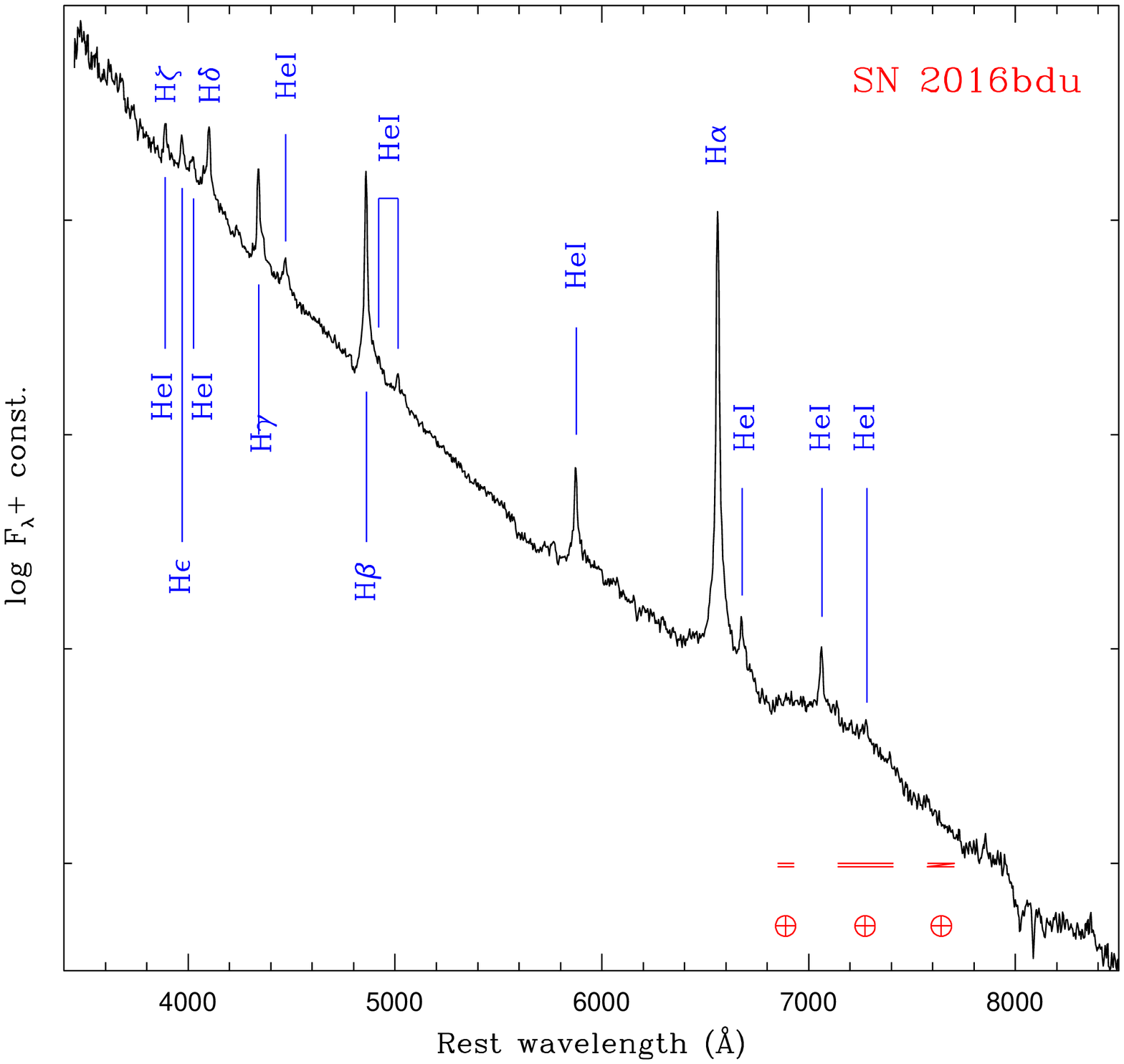}} 
    \caption{Top: Spectroscopic evolution of SN~2016bdu from around the epoch of the Event B maximum (JD = 2,457,541.5), to the early nebular phase (+78.9 d). Bottom: Identification of the strongest 
emission lines in the earliest SN~2016bdu spectrum. The regions contaminated by the strongest telluric bands are marked.} 
    \label{fig6} 
\end{figure}

Our extensive spectroscopic campaign for SN~2016bdu started on 2016 June 2, near the Event B light-curve peak. General information on the spectra is collected  
in Table \ref{tab:spectra}.\footnote{All of our spectra will be available in the WISeREP archive \protect\citep{yar12}.}
 Most of the spectra were taken at the parallactic angle \citep{fil82}, hence minimising 
differential flux losses.

The spectra were reduced with IRAF tasks\footnote{AFOSC and ALFOSC spectroscopic data were reduced using a dedicated graphic user interface  
developed by E. Cappellaro (http://sngroup.oapd.inaf.it/foscgui.html).}. One-dimensional spectra were first extracted from the two-dimensional  
frames, and then wavelength-calibrated using spectra of comparison lamps obtained with the same instrumental setup. The wavelength calibration was checked  
by measuring the positions of several night-sky lines, and (when necessary) shifted by a constant amount to match the expected wavelength of these  
lines. The flux calibration was performed using spectra of standard stars obtained during the same night as the SN observation, and the accuracy of the  
calibration was checked with the available coeval photometry; in cases of an overall flux discrepancy, a scaling factor was applied to calibrate the spectrum to the  
photometry. Finally, the strongest telluric absorption bands (in particular, O$_2$ and H$_2$O) were corrected using the spectra of the standard stars. 
 
\begin{figure} 
	\includegraphics[width=\columnwidth]{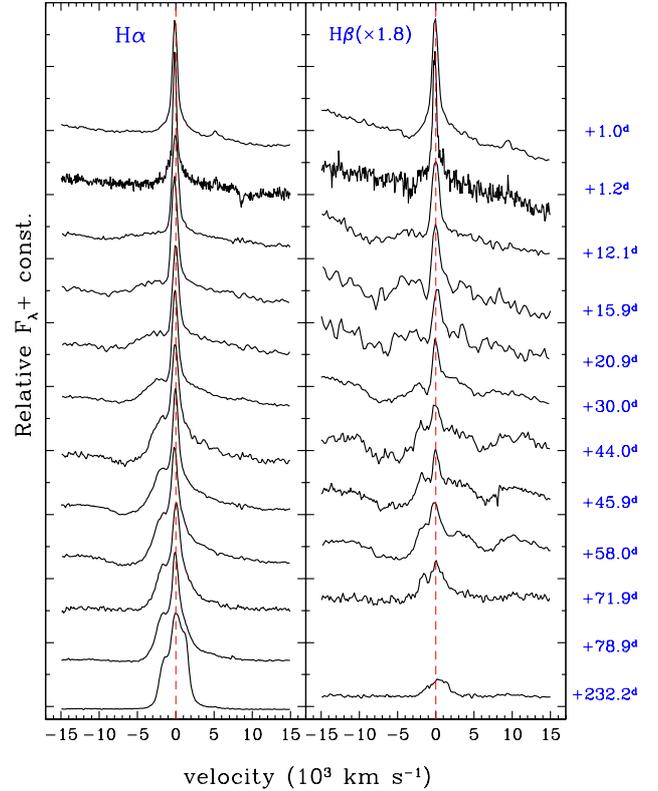} 
    \caption{Evolution of the H$\alpha$ (left panel) and H$\beta$ (right panel) line profiles in velocity space.  
The vertical red dashed lines mark the rest-velocity position of the two lines.} 
    \label{fig7} 
\end{figure}

The spectral sequence obtained  
during the first $\sim 80$ days after the Event B maximum is shown in the top panel of Figure \ref{fig6}, while line identifications in our earliest  
spectrum (phase +1 d) are reported in the bottom panel of Figure \ref{fig6}. The detailed evolution of the H$\alpha$ and H$\beta$ line profiles is shown  
in Figure \ref{fig7}.  
 
The early-time spectra resemble those of typical Type IIn SNe, and are characterised by a blue continuum 
(with a blackbody temperature $T_{\rm BB} = 17,000 \pm 1000$~K) and narrow emission lines of H and He~I. These lines show two velocity components: a narrow  
component with a full-width at half-maximum intensity (FWHM) velocity of $v_{\rm FWHM} = 400$ km s$^{-1}$ 
(as determined from the highest resolution FAST spectrum obtained on June 3) superposed on lower intensity, broader P~Cygni wings. 
In analogy with classical SNe~IIn, the narrow emission lines likely arise in a slow-moving, unshocked photoionised CSM, while the broader  
wings can be interpreted as being produced by electron scattering. The broad P~Cygni minimum of H$\beta$ is blueshifted by  $v = 3400$ km s$^{-1}$ (see Figure \ref{fig7}). 
This value is consistent with that derived from the position of the minimum of the He~I $\lambda$5876 absorption line.  
 
With time, the spectral continuum becomes redder, and the narrow components weaken relative to the broad emission lines. 
In particular, when the continuum has declined to $T_{\rm BB} = 13,500 \pm 1100$~K in the June 14 spectrum (phase +12.1 d), narrow He~I lines are no longer visible.  
The broad He~I $\lambda$5876 feature displays a P~Cygni profile whose velocity (from the position of the absorption minimum) is $v \approx 9000$ km s$^{-1}$.  
A similar broad P~Cygni component is also visible for the H lines, with H$\alpha$ having $v = 9500$ km s$^{-1}$ and H$\beta$ at $v = 8400$ km s$^{-1}$ (Figure \ref{fig7}), 
with blueshifted absorption wings extending up to 13,000 km s$^{-1}$. 
These velocities are likely representative of the fast-moving ejecta. The narrow H components are still visible, and exhibit 
a  P~Cygni profile, with the absorption component being blueshifted by about 1900 km s$^{-1}$ (as measured for 
the H$\beta$ line). 
 
A possible explanation for the evolution of the  H$\beta$ absorption trough (Figure \ref{fig7}) is the presence of 
a fast shell close to the SN, produced by a stellar wind with a mass-loss rate of 
$\sim 10^{-1}$ M$_\odot$ yr$^{-1}$.  
This interpretation, proposed by \citet{chu04} and \citet{des16} for the Type IIn SN~1994W, can also work for the early-time  
spectra of SN~2016bdu. 
The interaction between the massive SN ejecta with this wind produces a dense shell, which initially has a velocity of 3400 km s$^{-1}$ 
and, later on, slows down to 1900 km s$^{-1}$. 
 
On July 2 (phase 30 d), the spectrum of SN~2016bdu becomes redder ($T_{\rm BB} = 9300  \pm 1100$~K), and the broad absorption components of the 
Balmer lines are now more prominent. The narrow components of the H features are still visible, with the P~Cygni profiles blueshifted 
by about 1200 km s$^{-1}$. This velocity, which will stay roughly constant at later phases, is likely very close to the initial velocity of the shell. 
The He~I $\lambda$5876 line remains quite weak, and its very broad, boxy absorption profile suggests 
some contamination from a growing Na~I $\lambda\lambda$5890, 5896 doublet (Na~I~D). An alternative explanation for this line profile 
is an additional absorption contribution from an He-rich shell. 
At this epoch, the line velocities inferred 
for the broad H$\alpha$ and H$\beta$ absorptions are $v = 7300$ km s$^{-1}$ and $v = 6800$ km s$^{-1}$, respectively. Some Fe~II lines (in 
particular, those of the multiplet 42) are also barely detected.

Later on, from July 15 (phase 44 d), the now quite red ($T_{\rm BB} = 6100 \pm 800$~K) spectrum  develops relatively broad lines of metals in absorption, including several Fe~II multiplets,  
Na~I~D, and Ca~II. O~I $\lambda$7774 is also marginally detected. The Na~I~D feature now dominates over He I $\lambda$5876, while other He~I  
lines are no longer observed. 
 
On July 29 (58 d), we obtained a high-S/N spectrum which shows a blackbody temperature to similar the one at 44 d. 
The SN spectrum is still dominated by broad lines with P~Cygni profiles. Numerous metal lines 
are strong, including Fe~II (with average $v = 2800$ km s$^{-1}$), Na~I~D ($v = 3700$ km s$^{-1}$), Ca~II~H\&K, and the 
(weak) NIR triplet of Ca~II. The broad P~Cygni lines of the Balmer series are prominent, with the residual narrow 
H components being detectable only in H$\alpha$ and H$\beta$. The gas velocities, as estimated from the minima of the broad 
Balmer absorptions, are $v = 6700$ km s$^{-1}$ and $v = 5450$ km s$^{-1}$ for H$\alpha$ and H$\beta$, respectively (Figure \ref{fig7}). A prominent feature is also 
detected blueward of Na~I~D, at about 5600~\AA; we tentatively identify it as Sc~II. Finally, Ti~II lines very likely contribute to the blanketing at 
blue wavelengths. 
 
Late-time spectra were obtained on August 12 and August 19 (phases 71.9~d and  78.9~d, respectively).  
The spectrum has now significantly changed. The continuum is very weak, 
and the H lines are dominated by the broader component in emission. There are still detectable narrow components superposed on the broad lines. 
 The unblended He~I $\lambda$7065 feature becomes visible again (and relatively prominent), 
with $v_{\rm FWHM} = 2800$ km s$^{-1}$, along with a Na~I~D  plus He~I $\lambda$5876 blend, with a residual P~Cygni profile (the absorption mimimum is blueshifted by about 3800 km s$^{-1}$). 
A weak emission line at about 7200~\AA\ is probably the emerging [Ca~II] doublet (possibly blended with He I $\lambda$7281), and 
the NIR Ca~II triplet is now prominent. 
A broad and weak feature appearing at about 6300~\AA\ could be a signature of the growing [O~I]  $\lambda\lambda$6300, 6364 doublet. 
 
\begin{figure} 
	\includegraphics[width=\columnwidth]{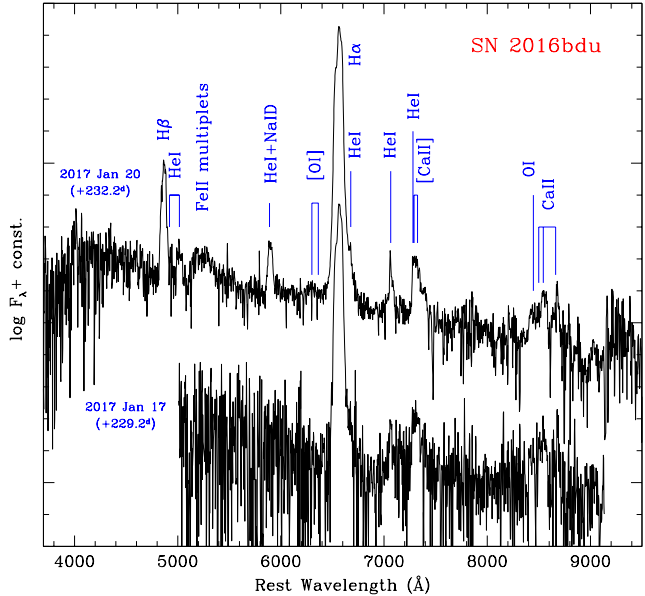} 
    \caption{Late-time ($\sim 200$--230 d) GTC+OSIRIS spectra of SN~2016bdu, along with identifications of the most prominent lines.} 
    \label{fig8} 
\end{figure} 
 
When the object was visible again in mid-January 2017 (phase $\sim 230$ d) after the seasonal gap, we obtained two additional spectra with GTC+OSIRIS (Figure \ref{fig8}). 
The spectra do not show any major changes from those obtained in August 2016, although the residual P~Cygni absorption features have now completely vanished. 
The overall H$\alpha$ line profile and the increased strength of He~I lines suggest that SN~2016bdu is still interacting with its CSM,  
although some broad features expected in SN spectra during the nebular phase are now clearly detected.  
The nebular lines (in particular [O~I], [Ca~II], and NIR Ca~II) are still much fainter than H$\alpha$. H$\beta$ is also detected, along with a number of He~I lines.  
Comprehensive line identifications for the two latest spectra are given in Figure \ref{fig8}.  
 
\begin{figure} 
	\includegraphics[width=\columnwidth]{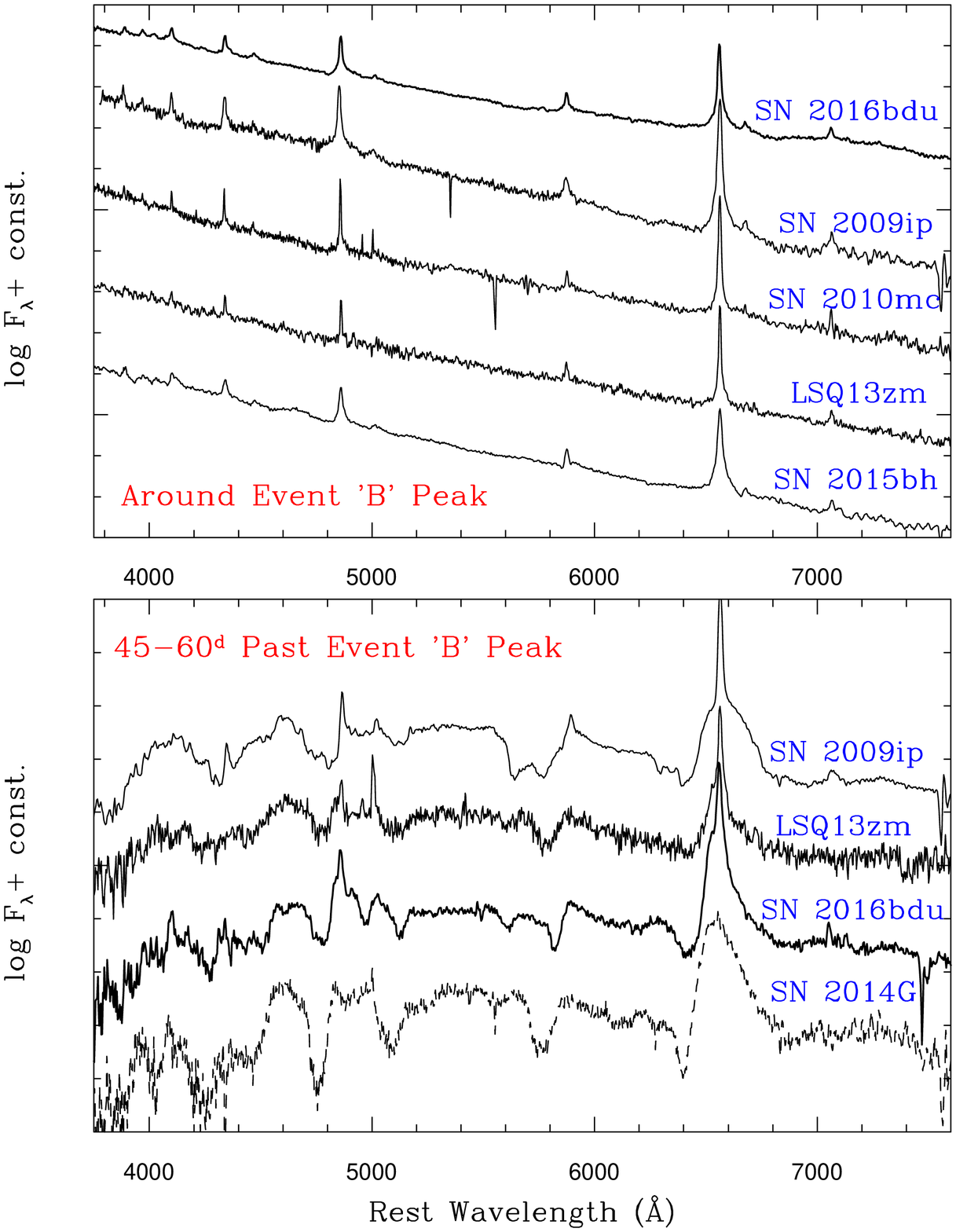} 
    \caption{Top: Comparison of spectra of SN~2009ip-like events obtained around the epoch of the Event B maximum \citep[the spectra are from][]{gra14,ofek13,leo16,ner16}.  
Bottom: Comparison of spectra of SNe 2009ip \citep{fra13b}, 
LSQ13zm \citep{leo16}, and 2016bdu with a spectrum of the more typical Type II SN~2014G \citep{giac16b}.} 
    \label{fig9} 
\end{figure} 
 
The H$\alpha$ FWHM in our latest spectra exceeds 3000 km s$^{-1}$, and the line has an asymmetric profile, with two shoulders:  
one blueshifted by about 1250 km s$^{-1}$, the other redshifted by nearly 900 km s$^{-1}$ (Figure \ref{fig7}). Although asymmetric Balmer  
line profiles are usually interpreted as signatures of asymmetric material ejection, the overall line shape of the H lines 
can be most likely explained as boxy profiles with a superimposed narrow P~Cygni component from the CSM. 
We also measure an H$\alpha$/H$\beta$ line ratio (Balmer decrement) of 11.  
Such large Balmer decrements are quite common in SNe~IIn \citep[e.g., SNe 1995G and 1996al;][]{pasto02,ben16}, and are frequently viewed as  
indicative of collisional excitation \citep{bra81}. An alternative explanation for the asymmetric Balmer line profiles and the large Balmer decrement 
could be that some dust is forming and is obscuring the receding material. However, this interpretation seems less plausible, as there is no evidence of a  
NIR excess in our late-time photometry. 
 
While the early-time spectra of SN 2016bdu (around the Event B maximum) are very similar to those of other 
SN~2009ip-like transients at the same phase (see Figure \ref{fig9}, top panel), later spectra of SN 2016bdu  
closely resemble those of typical SNe~II during the early H-envelope 
recombination phase (Figure \ref{fig9}, bottom panel). This supports the argument for a terminal SN explosion  for all SN 2009ip-like 
objects, although this claim is still debated since broad lines were also detected in pre-SN stages in some cases 
\citep[e.g.,][]{pasto13}. The different scenarios proposed for SN~2009ip, revised in the context of SN~2016bdu, will be discussed in Sect. \ref{discussion}. 
 
\section{Eruptions, mergers or SN explosions?} \label{discussion} 
 
As mentioned earlier, at least three SNe of this family have experienced a phase of major erratic variability lasting several years prior to the primary  
outburst, but the actual number is likely much larger. For example, in the more distant LSQ13zm and SN~2010mc, the earlier phase of erratic variability  
was probably below the detection threshold. All five objects experienced a major outburst, characterised by a light curve showing two brightening episodes.  
The duration and the luminosity of the two events are comparable in all SN~2009ip-like objects, which also show remarkable similarity in their spectral  
evolution (at least, at phases when spectroscopic observations exist; see Sect. \ref{spec}). This observed homogeneity is even more puzzling for  
progenitors possessing strongly asymmetric circumstellar environments \citep{lev14,mau14,ner16,tho16}, whose orientation with respect to the line of sight  
is expected to play an important role. The striking spectroscopic and photometric similarities of the five SN~2009ip-like transients suggest that these 
objects very likely arise from similar stars (or stellar systems) and may have undergone a comparable fate. 
 
The nature of SN~2009ip and similar objects has been widely discussed in the literature, and multiple scenarios have been offered, none of them 
supported by conclusive evidence. In this section we discuss the most plausible scenarios for these peculiar transients. 
 
\begin{enumerate} 
\item {\bf A major outburst (Event A) followed by shell-shell collision (Event B).} This scenario was first proposed by \citet{pasto13} for SN~2009ip. The argument was  
based on the evidence that broad spectral lines, with velocities comparable with those of real SNe, were observed during the 2008 to early-2012 impostor  
phase of SN~2009ip, long before the putative SN. In this view, Event A in July 2012, whose spectra showed broad P~Cygni lines, would be a huge outburst, with the subsequent Event B being the result of reprocessing of kinetic  
energy into radiation due to collision of the most recent mass ejection (Event A) with CSM collected during previous eruptive phases. 
This scenario is also supported by the lack of late-time nebular SN spectral signatures expected from the explosion of a massive star  
\citep{fra13b,mar14,gra14,fra15}. Based on energetic considerations, \citet{mor15} suggested that Event B in SN~2009ip was not caused by interaction 
of material expelled in a regular SN explosion with CSM, but rather from a shell-shell interaction. 
The mechanisms that trigger these major mass-loss events are debated, and plausible explanations invoke pulsational pair-instability \citep[e.g.,][]{woo07}, 
or interactions in a massive binary system \citep[e.g.,][]{kas10}. This latter scenario was also suggested for the impostor SN~2000ch \citep{pasto10}. 
 
An argument frequently used to rule out a terminal SN explosion is the absence of emission lines from nucleosynthetic byproducts in the late-time 
spectra of these objects \citep[see, e.g., the discussion in][for SN~2009ip]{fra15}. On the other hand, the late-time spectrum of the SN~2009ip-like  
event SN~2015bh exhibits prominent and relatively broad (about 3000 km s$^{-1}$) [Ca~II] $\lambda\lambda$7291, 7324 lines, even if the spectra still 
do not show unequivocal evidence of the broad [O~I] $\lambda\lambda$ 6300, 6364 features expected from a classical core-collapse SN. However, the [Ca~II] $\lambda\lambda$7291, 7324 emission 
is not necessarily a tracer of fresh nucleosynthetic elements, as it can be produced by primordial material. 
 
One of the key observational constraints for our understanding of SN~2009ip-like transients is the flattening of the late-time light curve 
observed in SN~2009ip, LSQ13zm, and SN~2015bh at phases later than 150--200 d. A flattening in the late-time light curve is also observed for SN~2016bdu. 
Although the spectral appearance in this phase suggests CSM interaction is still playing a key role, 
the exponential luminosity decline without significant variability may indicate that the progenitors have returned to a quiescent phase or eventually  
disappeared after core collapse. Although none of the above arguments is conclusive, the scenario invoking an outburst followed by shell-shell collisions appears to be less appealing.

\item {\bf A faint core-collapse SN (Event A) plus ejecta-CSM interaction (Event B).} This scenario for explaining the double-peaked light  
curve of the SN~2009ip-like transients was first proposed by \citet{mau13} for SN~2009ip, and expanded by \citet{smi14} who also suggested that the $\sim 60$~M$_\odot$  
progenitor was  a blue supergiant that exploded as a faint core-collapse SN during Event A. 
Such massive stars may, in fact, produce underenergetic ($10^{50}$ erg) explosions, and eject very little $^{56}$Ni owing to fallback onto the protoneutron star 
\citep{zam98,heg03,nom06,mor10,pej15,suk16}. The result is a weak core-collapse SN, consistent with the properties of Event A.  
A fallback SN (producing the Event A light curve) would also explain the weakness of the $\alpha$-element signatures in late-time spectra. 
The subsequent collision of the outer SN ejecta with the CSM from the more recent mass-loss events produces the major rebrightening during Event B \citep[see, e.g.,][]{chu91}. 
This scenario is also supported by a sequence of spectropolarimetric observations of SN~2009ip; these data suggest that SN~2009ip Event A is consistent with  
a prolate (possibly bipolar) SN explosion having a canonical kinetic energy ($E_{\rm K} \approx 10^{51}$~erg), and whose ejecta collide  
with an oblate CSM distribution \citep{mau14}.
 
\begin{figure} 
	\includegraphics[width=\linewidth]{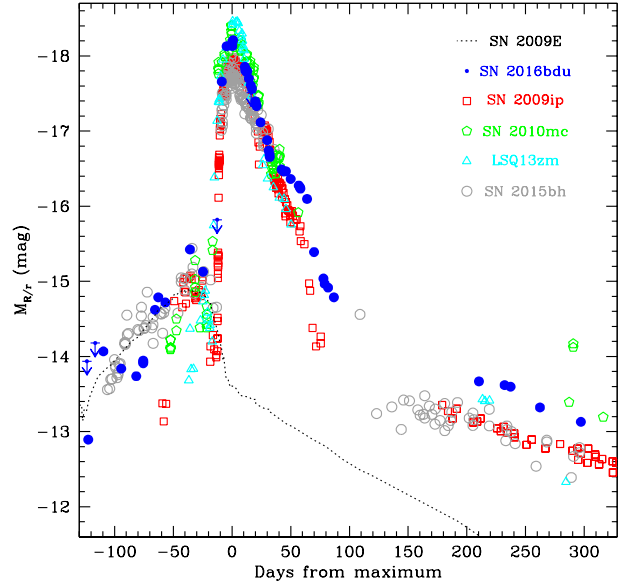} 
    \caption{Comparison of Johnson-Cousins $R$ (or Sloan-$r$) absolute light curves for our sample of SN~2009ip-like events with the light curve of the faintest known SN~1987A-like 
SN~2009E \citep{pasto12}. All data are on the Vega magnitude scale. The phases span from the maximum of Event B to the late-time $^{56}$Co decay tail. 
 The light curve of SN~2009E has been shifted arbitrarily in phase to approximately match the Event A peaks 
of SN~2009ip-like SNe, and has been dimmed by 1 mag.} 
    \label{fig10} 
\end{figure} 
 
A similar scenario has been invoked by \citet{ner16} for SN~2015bh based on the surprising similarity between the light curve of SN~2015bh during Event A and 
 weak SN~1987A-like events \citep[e.g., SN~2009E; see Figure \ref{fig10};][]{pasto12}. 
The light curve of SN~2016bdu during Event A is consistent with a faint core-collapse SN (perhaps SN~1987A-like; Figure \ref{fig10}). 
 
This scenario was not favoured by \citet{ofek13} for SN~2010mc on the basis of the following arguments. The impact of SN ejecta  
with the CSM generates collisionless shocks, which produce hard X-ray photons \citep{che12,svi12}. To reprocess these photons to visible light through  
Compton scattering, an optical depth on the order of a few times unity is necessary. Using the temperature and the emitting radius estimated from blackbody  
fits to the spectral continuum of SN~2010mc, \citet{ofek13} inferred that (for the assumed optical depths) the diffusion timescales were longer  
than the observed duration of the Event A peak of SN~2010mc. 
On the other hand, all the other transients of this group (including SN~2016bdu) have much longer rise times  
to the Event A peak than does SN~2010mc. In addition, an asymmetric gas distribution as observed in both SNe~2009ip and 2015bh  
\citep[e.g.,][]{lev14,mau14,ner16} may weaken the Ofek et al. arguments. 
  
\item {\bf An outburst (Event A) followed by an interacting SN explosion (Event B).} In this scenario, Event A would be the last and most energetic outburst  
of a sequence likely initiated a few years before (during the impostor phase). Then, the rise of Event B would be the direct signature of the true SN explosion.  
The blue spectra, characterised by prominent, narrow emission lines of H and He~I, would be produced in the CSM, initially photoionised by the SN shock breakout and  
later by the CSM-ejecta interaction. This scenario was first proposed for SN~2010mc \citep{ofek13} and a few other SNe~IIn in the PTF sample \citep{ofek14}. 
It was also suggested by \citet{leo16} to explain the very high velocities ($2.3 \times 10^4$ km s$^{-1}$)  
observed in spectra of LSQ13zm. In fact, when adopting for LSQ13zm the epoch of the initial rise of Event A as the time of the SN shock breakout,  
a photospheric velocity of more than $2 \times 10^4$ km s$^{-1}$ would be observed $\sim 70$~d after the SN explosion, which is unusual in typical  
core-collapse SNe. We also note that the spectra of LSQ13zm $\sim 50$ days after the Event B maximum are reminiscent of SN~II spectra during  
the photospheric phase (bottom panel of Figure \ref{fig9}).  
Moving up the epoch of core collapse to the onset of Event A (i.e., at least 3 months before the peak of Event B) would imply that  
the H-envelope recombination phase occurs rather late (4--5 months after the explosion).  
A similar sequence of events was also proposed for the pre-SN outburst of a Wolf-Rayet star two years before the explosion of the Type Ibn SN~2006jc  
\citep[][]{pasto07,fol07}, and the outburst of a putative super-asymptotic-giant-branch star was observed a few months before the explosion of the Type IIn-P SN 2011ht  
\citep[possibly an electron-capture SN;][]{fra13a,mau13b,smi13b}. However, other authors suggest that the properties of SNe~IIn-P are best explained as  
resulting from the interaction of subsequent shells produced by two nonterminal outbursts \citep{des09,hum12}. 
 
\item {\bf Binary interaction during Event A (and before) with a final merger (Event B).} 
 
The possibility that SN~2009ip-like events are produced in interacting binary systems has been previously mentioned. This would be consistent 
with pre-SN histories characterised by erratic variability and a multiple-shell CSM structure \citep{pasto13,mar14,gra14,mar15}. In addition, 
\citet{lev14} constrained the geometry of the SN~2009ip CSM to have an accretion disc (see also \citealt{mau14}). All of this suggests the presence of a companion star in a highly eccentric orbit, 
with interactions between the two stars during each periastron  passage triggering  ejections of material.  
 
\citet{kas13} proposed that Event B in SN~2009ip was powered by the accretion of several solar masses of gas onto the primary luminous blue variable (LBV) star. The secondary  
star may have survived the encounter or eventually merged onto the primary. This scenario has also been suggested by \citet{sok13} and \citet{sok16}  
for SNe~2009ip, 2010mc, and 2015bh \citep[see also][for a discussion of massive stellar mergers including LBVs]{san12,sel14,jus14,por16}.  
In these massive binary systems, formed by an LBV and a more compact companion, the high-velocity gas outflow seen in the spectra of both SN~2009ip  
and SN~2015bh during the impostor phase (hence months to years before Event A) would be powered by jets from the secondary star \citep{tse13}. 
\citet{gor16} proposed a hybrid explanation for SN~2015bh, with the core collapse of an evolved massive star while merging with a massive binary  
companion. In this context, we note that some bumps similar to those observed in SN~2009ip are also visible in the light curve of SN~2016bdu during  
the rise to the Event A maximum, and during the post-peak decline from Event B.  
 
Although binary interaction followed by a merging event cannot be definitely ruled out for SN~2016bdu, there are additional arguments that do not  
support this scenario. First, the spectra of the most promising merger candidates seem to evolve toward those of cool, M-type stars  
\citep[e.g.,][]{smi16b,bla17}, which is obviously not the case for SN~2009ip-like events. Second, all known stellar mergers follow clear correlations  
between the physical parameters of the progenitors and the luminosity of the outbursts \citep{koc14}. The known high-mass mergers also follow these  
correlations \citep[][and references therein]{mau17}, while the SN~2009ip-like transients are much more luminous for their expected stellar masses. 
 
\end{enumerate} 
 
In summary, SN~2016bdu is observationally a member of the SN~2009ip-like family. Although none of the alternative scenarios is ruled out, 
the spectroscopic features observed during the decline following Event B closely resemble those of rather normal SNe~II, favouring a 
terminal core-collapse explosion for SN~2016bdu (and, plausibly, for all members of this family). Whether core collapse happened 
at the onset of Event A or B is also controversial. Further clues supporting the terminal SN explosion for SN~2016bdu and all other clones of SN~2009ip 
will be provided from the comparison with the well-studied Type IIn SN~2005gl, below. 
 
\subsection{SN~2005gl: a link with SN~2009ip-like events?} \label{05gl} 
 
Important insights on the nature of SN~2016bdu may come from a comparison with the Type IIn SN 2005gl, an object sharing some similarity with 
SN~2009ip-like events, and for which a very luminous source at the position of the SN was identified in pre-SN {\it Hubble Space Telescope (HST)}  
archival images obtained in June 1997. This object was interpreted by \citet{gal07} as the probable SN progenitor. In particular, late-time observations  
of the site of SN~2005gl showed no trace of the putative progenitor \citep{gal09}, which provides strong support for the identification of the  
progenitor and direct evidence that a massive star, very likely an LBV, exploded to produce an SN~IIn. 
Identifying SN~2005gl as a member of the SN~2009ip-like family would provide reasonable arguments to claim that all members of this family experienced 
a similar fate. 
 
\begin{figure} 
	\includegraphics[width=\linewidth]{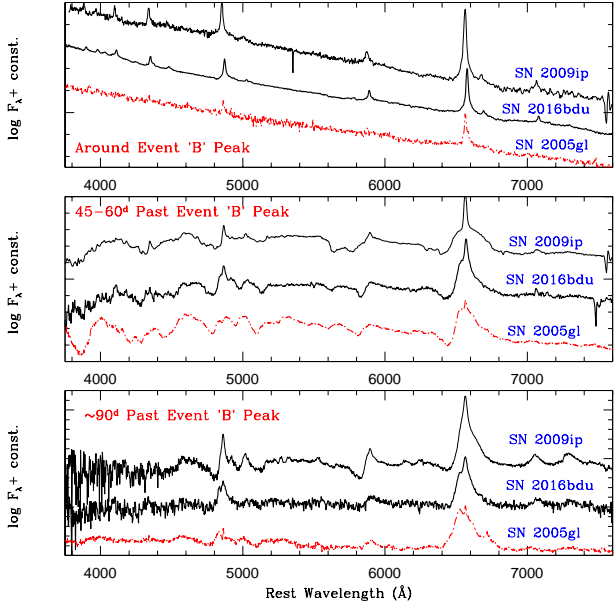} 
    \caption{Comparison of the spectra of SNe~2009ip, 2016bdu, and 2005gl obtained at three epochs around the Event B maximum (top), 1.5--2 months after the Event B maximum (middle),  
and about 3 months after the Event B maximum (bottom).  
The spectra of SNe 2005gl and 2009ip are taken from \citet{gal07,fra13b,gra14} and \citet{mar14}.} 
    \label{fig11} 
\end{figure} 
 
Here we provide a few observational arguments to support the similarity of SN 2005gl with members of the SN~2009ip group. In Figure \ref{fig11}, we  
compare spectra of SN~2005gl \citep{gal07} at three representative epochs with those of SNe~2009ip and 2016bdu at similar phases (indicated by labels  
in the Figure). The striking spectral similarity of SN~2005gl with the SN~2009ip-like events is evident both around maximum brightness and at later phases. The  
spectral evolution of these objects is not typical of SNe~IIn, whose spectra are always dominated by the narrow and intermediate-width emission lines,  
and never show broad P~Cygni profiles. 
 
We also compute an updated light curve of SN~2005gl from follow-up photometry obtained with the 0.76~m Katzman Automatic Imaging Telescope \citep[KAIT, at Lick Observatory;][]{fil01,leaman11}, along with new unfiltered observations from amateur astronomers 
calibrated to the $R$ band (details are given in Appendix \ref{app1b}, and the photometric measurements are reported in Table \ref{tab:2005gl}). 
 
\begin{figure*} 
	{\includegraphics[width=0.55\textwidth,angle=270]{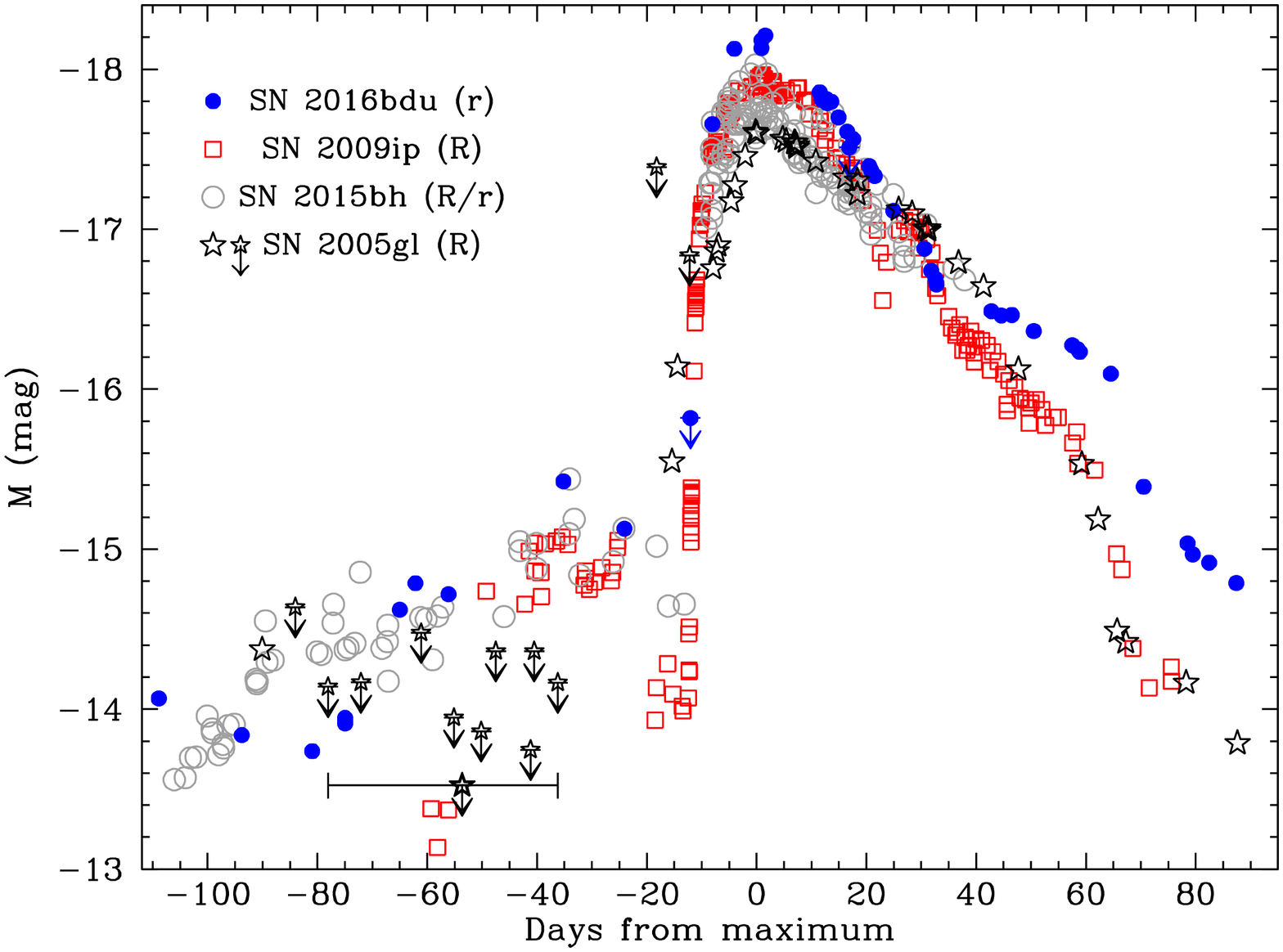} 
	\includegraphics[width=0.55\textwidth,angle=270]{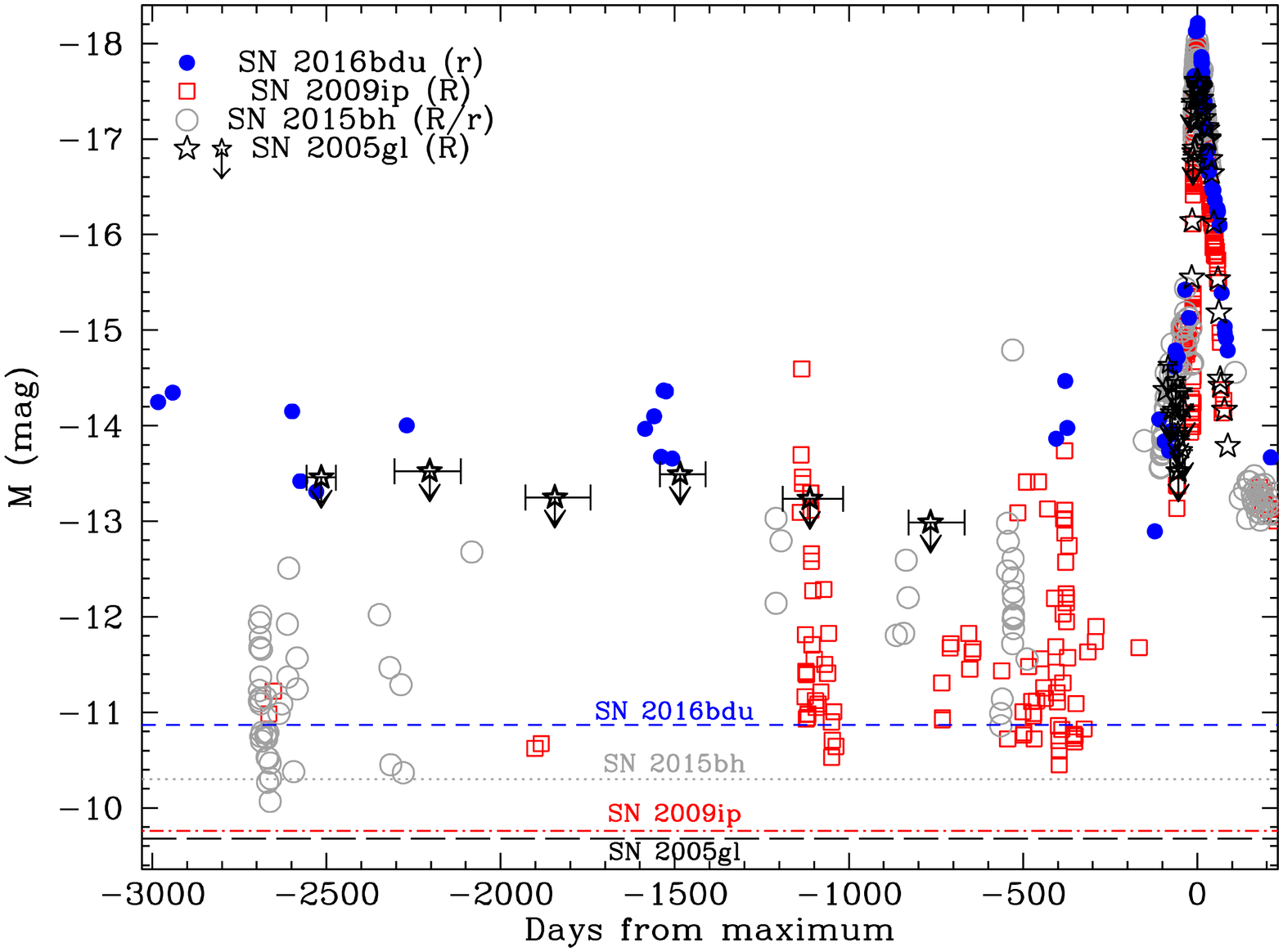}} 
    \caption{Top: comparison of the absolute Johnson-Cousins $R$-band and Sloan $r$-band light curves of SNe~2005gl, 2016bdu, 2009ip \citep{pasto13,fra13b},  
and 2015bh \citep{ner16} in the Vega magnitude system. The phase covered is that of the main outburst. Photometric points of SN~2005gl with horizontal error  
bars represent detection limits inferred from seasonal stacked images. Bottom: long-term absolute Johnson-Cousins $R$-band and/or Sloan $r$-band light curves  
of SNe~2005gl, 2016bdu, 2009ip \citep{pri13,pasto13,fra13b,mar14}, and 2015bh \citep{ner16}. While for SN~2005gl the detection limits are also reported, for  
all other events we show only real detections. The distance modulus of SN~2009ip is taken from \citet{fra13b}, while we revised that of SN~2005gl to $\mu = 34.11$ mag.  
For the line-of-sight extinction in the direction of SN~2005gl, we adopt the upper limit ($E(B-V) = 0.1$ mag) reported by \citet{gal07}. The faintest archival  
detections for the progenitors of SN~2016bdu (short-dashed blue line), SN~2009ip (dot-dashed red line), SN~2015bh (dotted grey line), and SN~2005gl (long-dashed  
black line) are also indicated. Except for SN~2016bdu, for which the weakest pre-outburst detection is in the Sloan $r$ band, all other pre-outburst sources 
were detected in archival {\it HST} images in $F547W$, $F555W$, or $F606W$, which can be approximated by the $V$ band.}
    \label{fig12} 
\end{figure*}

In the top panel of Figure \ref{fig12}, a close-up view of SN~2005gl and SNe~2009ip, 2015bh, and 2016bdu during their Events A/B is shown. 
A good match is also seen between the light curve of SN~2005gl and those of SN~2009ip-like transients during their Event B. 
SN~2005gl reaches an absolute peak magnitude $M_R \approx -17.6$, which is very close to (just marginally fainter than) that of our SN sample, 
In addition, the light curves of SN~2005gl and the SN~2009ip-like transients are quite similar, and evolve more rapidly than those of most SNe~IIn. 
 
Despite the similarity of the Event B light curves, there is no robust detection of an Event A in SN~2005gl down to an absolute magnitude of about  
$-13.7$. There is a single marginal detection with S/N = 2.7 ($M_R = -14.46$ mag) about 90 d  
before the light-curve maximum of SN~2005gl. It is unclear if this is a signature of a short-duration outburst, or a hot pixel accidentally  
located at the transient's position. The magnitude of this possible precursor is brighter than most of the detection limits measured in  
the subsequent $\sim 50$ d. 
 
In order to address whether SN~2005gl also had an erratic-variability impostor phase and a previously unseen Event A,  we analysed the KAIT images of the host galaxy of SN~2005gl prior to the outburst  (October 1998 to July 2005) using  
the template-subtraction technique, stacking the images to increase their depth. The template was obtained by median combining a large number of good-quality  
KAIT images obtained between 2007 and 2009, since we are confident there is no contamination from SN~2005gl or its progenitor at these epochs. In fact,  
\citet{gal09} found no signature of a source at the SN location in deep {\it HST} images obtained on September 2007, to a 2$\sigma$ limit of $V > 25.9$ mag.  
The long-term $R$-band light curve of SN~2005gl (including the deep detection limits obtained from seasonal stacked images) is compared with those of our  
sample of SN~2009ip-like transients during their impostor phases in the bottom panel of Figure \ref{fig12}. Unfortunately, after the template subtraction, there  
is no evidence of pre-SN outbursts brighter than $M_R \approx -13$ to $-13.5$~mag over this time period (Figure \ref{fig12}, bottom panel).  
While this limit is fainter than the variability observed for SN~2016bdu, many of the SN~2009ip-like events showed pre-Event~A variability at only 
moderately fainter luminosities; thus, it is possible that SN~2005gl did so as well.  
 
The surprising observational match of SN~2005gl with SN~2009ip-like transients has important implications on the interpretation of the nature of all these objects. 
The progenitor candidate identified by \citet{gal07} and \citet{gal09} had $V = 24.1 \pm 0.2$ mag, implying an absolute magnitude of $M_V \approx -9.7$.  
This source was extremely luminous, and was consistent with being an LBV \citep{gal09}. 
Whether this LBV was in a quiescent stage or in an eruptive phase is uncertain, as there was only one detection in a single band. 
 
We also inspected ground-based archival images of the field of SN~2016bdu, and found a source with $r = 23.46 \pm 0.39$ mag in very deep  
INT images taken on 1999 February 10. To our knowledge, this is the faintest detection available of a stellar source at the position of SN~2016bdu. 
Owing to the relatively low spatial resolution of the image, this source can be either 
the variable progenitor of SN~2016bdu, a blend of multiple objects, or even an unrelated background source. No detection was registered in Johnson-Bessell $V$ and Sloan $i$ images collected  
on 1998 June 19 to a limit of $V > 23.78$ and $i > 23.65$ mag.  
We note that the progenitors of other SN~2009ip-like events have been observed in archival {\it HST} images, and have absolute $V$-band magnitudes 
quite close to $-10$ mag \citep{smi10,fol11,ner16}. These faint progenitor detections are indicated with horizontal lines in Figure \ref{fig12}. 
 
Since post-explosion {\it HST} images indicate that the stellar source detected by \citet{gal07} at the position of SN~2005gl is now below the  {\it HST} 
detection threshold \citep{gal09}, this implies that the star finally exploded as a SN or, alternatively, the surviving star is  much fainter than the June 1997 detection.  
 
So far, we do not have unequivocal proof that any of the SN~2009ip-like transients are associated with terminal SN explosions. Although the spectra obtained during Event A, and  
at 1--3 months after the maximum of Event B, show features closely resembling those of typical SNe~II, late-time follow-up observations of SN~2009ip (over 3~yr  
after the peak of Event B) show only marginal evidence of the classical nebular features of a core-collapse SN   
\citep[e.g., the prominent forbidden lines of O~I and Ca~II expected in the explosion of massive stars; see][and references therein]{fra15}. However, the late-time photometric 
evolution of this object (but also, for example, that of SN~2015bh) is slower than the expected decay rate of $^{56}$Co,  
without any signature of the photometric fluctuations observed during past evolutionary stages. This evolution is consistent with that of a genuine SN interacting 
with its CSM. More importantly, SN~2009ip is still fading, and it now has a magnitude comparable with that of the pre-explosion progenitor \citep{smi16}, making the terminal  
explosion scenario plausible for this event. If future observations demonstrate the disappearance of SN~2009ip-like transients, their similarity with SN~2005gl  
along with the observational clues mentioned above provide a robust argument for the death of their progenitor stars, as suggested by 
several authors \citep{mau13,mau14,smi14,leo16,ner16}. 
 
\section{Summary} \label{concl} 
 
In this paper, we presented optical and NIR observations of the recent Type IIn SN~2016bdu, along with additional optical photometric data for SN~2005gl.  
There is a striking observational match among SN~2016bdu, SN~2009ip, and other similar transients, as follows. 
\begin{itemize} 
\item Most have a pre-SN impostor phase characterised by an erratically variable light curve, similar to what is currently observed for the impostor SN~2000ch \citep{wag04,pasto10}. 
\item They have two sequential luminous outbursts (Events A and B) with similar structures. 
\item There is a dramatic decline of the luminosity after the maximum of Event B, followed by a flattening of 
the late-time light curve to a decline rate slower than that expected from $^{56}$Co decay into $^{56}$Fe.  
\end{itemize} 
 
Although we have no spectra of SN~2016bdu during Event A, we note that the spectroscopic evolution during Event B is  
remarkably homogeneous as well:  
\begin{itemize} 
\item At maximum brightness, their spectra are blue and show narrow emission lines of H and He~I, similar to those observed in the spectra of young SNe~IIn. 
\item During the steep post-peak decline (1--2 months after maximum), the spectra develop broad P~Cygni lines of H and metals, becoming similar  
to those of noninteracting (or mildly interacting) Type II SNe. 
\item Finally, at late phases (when the light curve flattens), the dominant spectral features are intermediate-width Balmer emission lines, with asymmetric profiles, 
likely indicative of enhanced ejecta-CSM interaction. The asymmetric H$\alpha$ line profile may also indicate asymmetries in the gas distribution or,  
less likely, dust formation. 
\end{itemize} 
 
We also note the striking similarity of SN~2016bdu and other SN~2009ip-like transients with the Type IIn SN 2005gl,  
the first interacting SN whose LBV progenitor was identified in pre-explosion images and not recovered in post-explosion images \citep{gal09},  
indicating that the star had finally exploded as a SN. 
 
Late-time observations might provide new insights into the nature of SN~2016bdu and similar objects through the detection of 
spectral signatures of a core-collapse SN remnant, or the disappearance of their progenitor stars. 
Nonetheless, the surprising similarity of the spectra after the Event B maximum with those of more typical SNe~II, and the excellent observational  
match with the genuine SN~IIn 2005gl, provide support for the terminal SN explosion scenario for SN~2016bdu and all members of the SN~2009ip-like family. 
 
\section*{Acknowledgements} 
 
We acknowledge S. Ascenzi, S. B. Cenko, R. Kotak, P. E. Nugent, N. Primak, A.~S.~B. Schultz, D.~E. Wright, S. Yang for their contribution to the observations 
and useful suggestions, and Y. Cai for reading the paper before submission. 
A.P., N.E.R., S.B., and G.T. are partially supported by the PRIN-INAF 2014 with the project ``Transient Universe: unveiling new types of stellar explosions with PESSTO.'' 
N.E.R. acknowledges financial support by MIUR PRIN 2010-2011, ``The dark Universe and the cosmic evolution of baryons: from current surveys to Euclid.'' 
M.F. acknowledges the support of a Royal Society -- Science Foundation 
Ireland University Research Fellowship. This work is partly supported 
by the European Union FP7 programme through ERC grant number 320360. 
K.Z.S. and C.S.K. are supported by US National Science Foundation (NSF) grants AST-1515876 and AST-1515927.  
S.D. is supported by Project 11573003 supported by NSFC and the ``Strategic Priority Research Program - The 
Emergence of Cosmological Structures'' of the Chinese Academy of Sciences 
(Grant No. XDB09000000). B.J.S. is supported by 
the National Aeronautics and Space Administration (NASA) 
through Hubble Fellowship grant HST-HF-51348.001 awarded  
by the Space Telescope Science Institute (STScI), 
which is operated by the Association of Universities for Research in 
Astronomy, Inc. (AURA) under NASA contract NAS 5-26555. 
A.V.F. and W.Z. are grateful for financial 
assistance from NSF grant AST-1211916, the TABASGO 
Foundation, and the Christopher R. Redlich Fund; they also acknowledge 
support through grant HST-AR-14295 from STScI, 
which is operated by AURA under NASA contract NAS 5-26555. 
The work of A.V.F. was completed at the Aspen Center for Physics, 
which is supported by NSF grant PHY-1607611; he thanks the Center for 
its hospitality during the neutron stars workshop in June and July 2017. 
T.W.-S.H. is supported 
by the DOE Computational Science Graduate Fellowship, grant number 
DE-FG02-97ER25308. 
Support for J.L.P. is in part by FONDECYT through the grant 1151445 and by the Ministry of Economy, Development, and Tourisms Millennium Science Initiative  
through grant IC120009, awarded to The Millennium Institute of Astrophysics, MAS. 
S.J.S. acknowledges (FP7/2007-2013)/ERC grant 291222.  
T.K. acknowledges financial support by the Emil Aaltonen Foundation. 
J.H. acknowledges financial support from the Finnish Cultural Foundation and the Vilho, Yrj\"o and Kalle V\"ais\"al\"a Foundation of the Finnish Academy of Science and Letters. 
A.G.-Y. is supported by the EU/FP7 via ERC grant No. 307260, the 
Quantum Universe I-Core program by the Israeli Committee 
for Planning and Budgeting, and the ISF; and by a Kimmel 
award. 
E.O.O. is grateful to support by grants  Israel Science Foundation, Minerva, and the I-CORE Program of the Planning and Budgeting Committee and The Israel Science Foundation. 
M.S. acknowledges generous funding from the Danish Agency for Science and Technology and Innovation realised through a Sapere Aude Level II grant, and from the Villum foundation.  
 
The CRTS survey is supported by NSF grants AST-1313422 and AST-1413600. 
ATLAS observations were supported by NASA grant NN12AR55G. 
NUTS is supported in part by the Instrument Center for Danish Astrophysics (IDA). 
We thank Las Cumbres Observatory and its staff for their continued support 
of ASAS-SN. ASAS-SN is funded in part by the Gordon and Betty Moore 
Foundation through grant GBMF5490 to the Ohio State University, NSF grant 
AST-1515927, the Center for Cosmology and AstroParticle Physics (CCAPP) at 
OSU, the Chinese Academy of Sciences South America Center for Astronomy 
(CASSACA), the Mt. Cuba Astronomical Foundation, and George Skestos. 
 
The Pan-STARRS1 Surveys (PS1) have been made possible through contributions of the Institute for Astronomy, the University of Hawaii, the Pan-STARRS Project Office, the Max-Planck Society and its participating institutes, the Max Planck Institute for Astronomy, Heidelberg and the Max Planck Institute for Extraterrestrial Physics, Garching, The Johns Hopkins University, Durham University, the University of Edinburgh, Queen's University Belfast, the Harvard-Smithsonian Center for Astrophysics, the Las Cumbres Observatory Global Telescope Network Incorporated, the National Central University of Taiwan, STScI, NASA under Grant No. NNX08AR22G issued through the Planetary Science Division of the NASA Science Mission Directorate, the US NSF under Grant No. AST-1238877, the University of Maryland, and Eotvos Lorand University (ELTE). Operation of the Pan-STARRS1 telescope is supported by NASA under Grant No. NNX12AR65G and Grant No. NNX14AM74G issued through the NEO Observation Program.  
 
This work is based in part on observations made with the Nordic Optical Telescope (NOT), operated on the island of La Palma jointly by Denmark, Finland, Iceland, 
Norway, and Sweden, in the Spanish Observatorio del Roque de los Muchachos of the Instituto de Astrof\'isica de Canarias; 
the 1.82~m Copernico Telescope of INAF-Asiago Observatory; the Gran Telescopio Canarias (GTC), installed in the Spanish Observatorio del Roque  
de los Muchachos of the Instituto de Astrof\'isica de Canarias, in the Island of La Palma; the Tillinghast Telescope of the Fred Lawrence Whipple Observatory; 
the Iowa Robotic Telescope which is located at the Winer Observatory (Arizona) and is scheduled and operated remotely from the University of Iowa; 
the Catalina Real Time Survey (CRTS) Catalina Sky Survey (CSS) 0.7~m Schmidt Telescope; and 
the Liverpool Telescope operated on the island of La Palma by Liverpool John Moores University at the Spanish Observatorio del Roque de los Muchachos of the Instituto de Astrofisica de Canarias with financial support from the UK Science and Technology Facilities Council. 
KAIT and its ongoing operation were made possible by donations from 
Sun Microsystems, Inc., the Hewlett-Packard Company, 
AutoScope Corporation, Lick Observatory, the NSF, the University of 
California, the Sylvia \& Jim Katzman Foundation, and the TABASGO 
Foundation.  Research at Lick Observatory is partially supported by a 
generous gift from Google. 
 
This work made use of the NASA/IPAC Extragalactic Database (NED), which is operated by the Jet Propulsion Laboratory, California Institute of Technology, under contract 
with NASA. We also used NASA's Astrophysics Data System. 
This publication made also use of data products from the Two Micron All Sky Survey, which is a joint project of the University of Massachusetts and the Infrared Processing and Analysis Center/California Institute of Technology, funded by NASA and the NSF. 
 
 

 
 
\appendix 
 
\section{Photometric data for SN~2016bdu} \label{app1a} 
 
Johnson-Bessell $B$ and $V$, Sloan $u$, $g$, $r$, $i$, and $z$, and NIR $J$, $H$, and $K$ magnitudes of SN~2016bdu are reported in Tables \ref{tab:2016bdu_JB}, \ref{tab:2016bdu_SDSS}, and \ref{tab:2016bdu_NIR}, respectively. 
 

\begin{table*} 
	\centering 
	\caption{Optical photometry (Johnson-Bessell, Vega mag) of SN~2016bdu.} 
	\label{tab:2016bdu_JB} 
	\begin{tabular}{CCCCc}  
		\hline 
		Date & JD & $B$ & $V$ & Instrument\\ 
		\hline 
1998-04-23 &  2450926.89 &         --       &    $>$19.84      &   1 \\ 
1998-06-19 &  2450984.43 &         --       &    $>$23.78      &   2 \\ 
2001-07-01 &  2452091.69 &         --       &    $>$20.94      &   1 \\ 
2002-03-02 &  2452336.05 &         --       &    $>$20.40      &   1 \\ 
2002-05-21 &  2452395.72 &         --       &    $>$21.21      &   1 \\  
2003-02-19 &  2452690.05 &         --       &    $>$20.45      &   1 \\ 
2016-02-18 &  2457436.91 &         --       &    $>$17.64      &   3 \\         
2016-02-19 &  2457438.14 &         --       &    $>$18.37      &   3 \\          
2016-02-20 &  2457438.91 &         --       &    $>$17.27      &   3 \\          
2016-02-22 &  2457440.98 &         --       &    $>$17.15      &   3 \\          
2016-02-24 &  2457443.17 &         --       &    $>$16.01      &   3 \\          
2016-02-25 &  2457443.98 &         --       &    $>$17.44      &   3 \\          
2016-02-27 &  2457446.11 &         --       &    $>$17.79      &   3 \\          
2016-02-29 &  2457447.97 &         --       &    $>$18.12      &   3 \\          
2016-03-02 &  2457450.01 &         --       &    $>$18.21      &   3 \\          
2016-03-04 &  2457451.99 &         --       &    $>$18.42      &   3 \\          
2016-03-07 &  2457454.87 &         --       &    $>$18.39      &   3 \\          
2016-03-11 &  2457458.93 &         --       &    $>$18.51      &   3 \\          
2016-03-14 &  2457461.92 &         --       &    $>$17.75      &   3 \\          
2016-03-16 &  2457463.86 &         --       &    $>$17.35      &   3 \\          
2016-03-19 &  2457466.84 &         --       &    $>$17.29      &   3 \\          
2016-03-22 &  2457469.99 &         --       &    $>$16.93      &   3 \\          
2016-03-25 &  2457472.95 &         --       &    $>$17.04      &   3 \\          
2016-03-28 &  2457475.85 &         --       &    $>$18.06      &   3 \\          
2016-03-31 &  2457478.89 &         --       &    $>$18.38      &   3 \\          
2016-04-03 &  2457481.97 &         --       &    $>$18.31      &   3 \\          
2016-04-10 &  2457488.85 &         --       &    $>$18.24      &   3 \\          
2016-04-12 &  2457490.97 &         --       &    $>$18.42      &   3 \\          
2016-04-16 &  2457494.77 &         --       &    $>$17.35      &   3 \\          
2016-04-20 &  2457498.96 &         --       &    $>$16.83      &   3 \\          
2016-04-23 &  2457501.89 &         --       &    $>$17.20      &   3 \\          
2016-04-27 &  2457505.91 &         --       &    $>$18.12      &   3 \\          
2016-04-28 &  2457506.83 &         --       &    $>$18.40      &   3 \\          
2016-05-02 &  2457510.89 &         --       &    $>$18.47      &   3 \\          
2016-05-05 &  2457513.89 &         --       &    $>$18.72      &   3 \\          
2016-05-05 &  2457513.90 &         --       &    $>$19.32      &   4 \\ 
2016-05-13 &  2457521.78 &         --       &    $>$17.68      &   3 \\          
2016-05-15 &  2457523.93 &         --       &    $>$17.77      &   3 \\         
2016-05-20 &  2457528.74 &         --       &    $>$17.33      &   3 \\          
2016-05-21 &  2457529.89 &         --       &    $>$17.14      &   3 \\          
2016-05-29 &  2457537.89 &         --       &    16.509 (0.084) &   3 \\          
2016-05-30 &  2457538.80 &         --       &    16.486 (0.075) &   3 \\          
2016-06-02 &  2457541.81 &  16.529 (0.019)  &    16.478 (0.036) &   5 \\ 
2016-06-03 &  2457542.85 &         --       &    16.343 (0.033) &   4 \\ 
2016-06-03 &  2457542.88 &         --       &    16.556 (0.132) &   3 \\          
2016-06-04 &  2457543.88 &         --       &    16.426 (0.084) &   3 \\          
2016-06-07 &  2457546.85 &         --       &    16.500 (0.024) &   4 \\ 
2016-06-09 &  2457548.87 &         --       &    16.606 (0.084) &   3 \\          
2016-06-12 &  2457551.79 &         --       &    16.909 (0.143) &   3 \\ 
2016-06-13 &  2457552.80 &  16.900 (0.049)  &    16.774 (0.065) &   6 \\ 
2016-06-14 &  2457553.56 &  16.961 (0.036)  &    16.807 (0.022) &   7 \\ 
2016-06-14 &  2457553.80 &  16.994 (0.065)  &    16.791 (0.074) &   6 \\ 
2016-06-16 &  2457555.85 &  17.119 (0.070)  &    16.858 (0.107) &   6 \\ 
2016-06-17 &  2457556.83 &         --       &    16.805 (0.141) &   6 \\ 
2016-06-17 &  2457557.41 &  17.201 (0.209)  &         --        &   8 \\ 
2016-06-20 &  2457559.82 &         --       &    17.046 (0.271) &   3 \\     
2016-06-22 &  2457561.92 &  17.444 (0.030)  &    17.156 (0.041) &   8 \\ 
2016-06-23 &  2457562.79 &         --       &    17.45  (0.16)  &   3 \\ 
2016-06-29 &  2457568.78 &         --       &    $>$18.06       &   3 \\ 
2016-07-03 &  2457572.73 &         --       &    17.959 (0.106) &   5 \\     
2016-07-04 &  2457573.75 &         --       &    17.974 (0.099) &   5 \\ 
2016-07-04 &  2457574.50 &  18.619 (0.024)  &    18.054 (0.032) &   7 \\ 
\hline                                                                                       
\end{tabular}                                                                               
\end{table*} 
\begin{table*} 
\contcaption{} 
	\begin{tabular}{CCCCc}  
		\hline 
		Date & JD & $B$ & $V$ & Instrument\\ 
		\hline 
2016-07-13 &  2457582.78 &         --       &    18.23  (0.15)  &   4 \\ 
2016-07-15 &  2457584.67 &  19.084 (0.134)  &         --        &   5 \\ 
2016-07-15 &  2457585.48 &  19.101 (0.032)  &    18.348 (0.030) &   7 \\ 
2016-07-21 &  2457591.40 &  19.258 (0.022)  &    18.464 (0.020) &   7 \\ 
2016-07-28 &  2457598.42 &  19.415 (0.027)  &    18.578 (0.029) &   7 \\ 
2016-08-04 &  2457605.41 &  20.075 (0.036)  &    19.067 (0.023) &   7 \\ 
2016-08-06 &  2457606.79 &         --       &    $>$19.19       &   4 \\ 
2016-08-12 &  2457613.44 &  20.781 (0.078)  &    19.930 (0.050) &   9 \\ 
2016-08-18 &  2457619.39 &  21.115 (0.088)  &    20.205 (0.030) &   7 \\ 
2016-08-23 &  2457624.31 &     --           &    20.341 (0.143) &   8 \\ 
2016-08-27 &  2457628.39 &  21.316 (0.066)  &    20.419 (0.057) &   7 \\ 
2016-12-23 &  2457746.16 &     --           &    $>$19.88       &   5 \\ 
2016-12-29 &  2457751.71 &     --           &    21.952 (0.101) &   7 \\ 
2016-12-31 &  2457746.16 &     --           &    $>$19.73       &   5 \\ 
2017-01-05 &  2457758.78 &  22.836 (0.299)  &     --            &   7 \\ 
2017-01-24 &  2457777.53 &  $>$22.87        &    22.227 (0.581) &   8 \\ 
2017-02-19 &  2457803.73 &  23.605 (0.396)  &    22.417 (0.241) &   7 \\ 
2017-03-26 &  2457838.61 &     --           &    23.012 (0.129) &   7 \\ 
		\hline 
	\end{tabular} 
 
\begin{flushleft} 
 1 = NEAT images; 2 = 2.54~m INT + WFC; 3 = ASAS-SN telescopes; 4 = ATLAS telescopes; 5 = Meade 10'' LX-200 Telescope + Apogee Alte F-47 CCD;  6 = 0.51~m Iowa Robotic Telescope + Apogee Alta F-47 CCD; 7 = 2.56~m NOT + ALFOSC; 8 = 1.82~m Copernico Telescope + AFOSC; 9 = 3.54~m TNG + LRS. 
\end{flushleft} 
\end{table*}

\begin{table*} 
	\centering 
	\caption{Optical photometry (Sloan AB mag) of SN~2016bdu.} 
	\label{tab:2016bdu_SDSS} 
	\begin{tabular}{CCCCCCCc}  
		\hline 
		Date  & JD & $u$ & $g$ & $r$ & $i$ & $z$ & Instrument\\ 
		\hline 
1998-06-19 &  2450984.44  &      --        &       --        &       --         &  $>$23.65     &       --        &  1 \\ 
1999-02-10 &  2451219.72  &      --        &       --        &  23.463 (0.391)  &       --      &       --        &  1 \\ 
2003-04-01 &  2452730.79  &      --        &       --        &  $>$20.00$^\oplus$&       --      &       --        &  2 \\ 
2004-03-19 &  2453083.88  &      --        &       --        &  $>$20.74$^\oplus$&       --      &       --        &  2 \\ 
2004-04-19 &  2453114.80  &      --        &       --        &  $>$20.79$^\dag$  &       --      &       --        &  2 \\ 
2004-04-28 &  2453123.78  &      --        &       --        &  $>$19.93$^\dag$  &       --      &       --        &  2 \\ 
2004-05-20 &  2453145.89  &      --        &       --        &  $>$20.69$^\dag$  &       --      &       --        &  2 \\ 
2005-03-08 &  2453437.42  &      --        &       --        &  $>$20.63        &       --      &       --        &  2 \\ 
2005-03-14 &  2453443.43  &      --        &       --        &  20.60 (0.37)    &       --      &       --        &  2 \\ 
2005-03-31 &  2453460.39  &      --        &       --        &   $>$20.64       &       --      &       --        &  2 \\ 
2005-04-11 &  2453471.32  &      --        &       --        &   $>$20.62       &       --      &       --        &  2 \\ 
2005-04-19 &  2453479.20  &      --        &       --        &   $>$20.39       &       --      &       --        &  2 \\ 
2005-05-01 &  2453491.22  &      --        &       --        &   $>$20.72       &       --      &       --        &  2 \\ 
2005-05-14 &  2453504.27  &      --        &       --        &  20.65 (0.62)    &       --      &       --        &  2 \\ 
2005-06-06 &  2453527.29  &      --        &       --        &   $>$20.33       &       --      &       --        &  2 \\ 
2006-02-04 &  2453771.52  &      --        &       --        &   $>$20.69       &       --      &       --        &  2 \\ 
2006-03-03 &  2453798.47  &      --        &       --        &   $>$20.39       &       --      &       --        &  2 \\ 
2006-03-29 &  2453824.57  &      --        &       --        &   $>$20.60       &       --      &       --        &  2 \\ 
2006-04-08 &  2453834.11  &      --        &       --        &   $>$20.27       &       --      &       --        &  2 \\ 
2006-04-27 &  2453852.99  &      --        &       --        &   $>$20.26       &       --      &       --        &  2 \\ 
2006-05-06 &  2453862.25  &      --        &       --        &   $>$20.49       &       --      &       --        &  2 \\ 
2006-06-01 &  2453888.17  &      --        &       --        &   $>$20.49       &       --      &       --        &  2 \\ 
2006-06-16 &  2453902.82  &      --        &       --        &   $>$18.75       &       --      &       --        &  2 \\ 
2007-02-10 &  2454141.96  &      --        &       --        &   $>$20.11       &       --      &       --        &  2 \\ 
2007-02-25 &  2454156.80  &      --        &       --        &   $>$20.66       &       --      &       --        &  2 \\ 
2007-03-10 &  2454169.81  &      --        &       --        &   $>$21.10       &       --      &       --        &  2 \\ 
2007-03-17 &  2454176.91  &      --        &       --        &   $>$20.45       &       --      &       --        &  2 \\ 
2007-04-19 &  2454209.83  &      --        &       --        &   $>$20.69       &       --      &       --        &  2 \\ 
2007-04-26 &  2454216.66  &      --        &       --        &   $>$20.42       &       --      &       --        &  2 \\ 
2007-05-15 &  2454235.72  &      --        &       --        &   $>$20.58       &       --      &       --        &  2 \\ 
2007-05-24 &  2454244.74  &      --        &       --        &   $>$20.28       &       --      &       --        &  2 \\ 
2007-06-05 &  2454256.75  &      --        &       --        &   $>$20.63       &       --      &       --        &  2 \\ 
2007-06-12 &  2454263.74  &      --        &       --        &   $>$20.31       &       --      &       --        &  2 \\ 
2007-06-21 &  2454272.69  &      --        &       --        &  20.44 (0.67)    &       --      &       --        &  2 \\ 
2008-01-17 &  2454483.01  &      --        &       --        &   $>$20.51       &       --      &       --        &  2 \\ 
2008-02-12 &  2454508.83  &      --        &       --        &   $>$20.56       &       --      &       --        &  2 \\ 
2008-02-18 &  2454514.87  &      --        &       --        &   $>$20.38       &       --      &       --        &  2 \\ 
2008-03-02 &  2454527.86  &      --        &       --        &   $>$20.48       &       --      &       --        &  2 \\ 
2008-03-08 &  2454533.85  &      --        &       --        &   $>$20.59       &       --      &       --        &  2 \\ 
2008-03-24 &  2454549.92  &      --        &       --        &   $>$20.46       &       --      &       --        &  2 \\ 
2008-04-01 &  2454557.86  &      --        &       --        &  20.32 (0.60)    &       --      &       --        &  2 \\ 
2008-04-08 &  2454564.83  &      --        &       --        &   $>$20.56       &       --      &       --        &  2 \\ 
2008-04-14 &  2454570.69  &      --        &       --        &   $>$20.42       &       --      &       --        &  2 \\ 
2008-05-03 &  2454589.75  &      --        &       --        &   $>$20.52       &       --      &       --        &  2 \\ 
2008-05-13 &  2454599.64  &      --        &       --        &  20.22 (0.68)    &       --      &       --        &  2 \\ 
2008-06-12 &  2454629.74  &      --        &       --        &   $>$20.53       &       --      &       --        &  2 \\ 
2008-12-07 &  2454807.98  &      --        &       --        &   $>$20.74       &       --      &       --        &  2 \\ 
2009-02-26 &  2454888.88  &      --        &       --        &   $>$20.63       &       --      &       --        &  2 \\ 
2009-03-17 &  2454907.96  &      --        &       --        &   $>$20.55       &       --      &       --        &  2 \\ 
2009-03-24 &  2454914.81  &      --        &       --        &   $>$20.68       &       --      &       --        &  2 \\ 
2009-04-02 &  2454923.71  &      --        &       --        &   $>$20.53       &       --      &       --        &  2 \\ 
2009-04-21 &  2454942.74  &      --        &       --        &  20.41 (0.58)    &       --      &       --        &  2 \\ 
2009-05-13 &  2454964.27  &      --        &       --        &   $>$20.90       &       --      &       --        &  3 \\ 
2009-05-15 &  2454966.27  &      --        &       --        &  21.15 (0.27)    &       --      &       --        &  3 \\ 
2009-05-17 &  2454968.79  &      --        &       --        &   $>$20.59       &       --      &       --        &  2 \\ 
2009-05-19 &  2454970.28  &      --        &       --        &   $>$19.48       &       --      &       --        &  3 \\ 
2009-05-30 &  2454981.70  &      --        &       --        &   $>$20.40       &       --      &       --        &  2 \\ 
2009-06-30 &  2455012.20  &      --        &       --        &  21.26 (0.35)    &       --      &       --        &  3 \\ 
2009-07-28 &  2455040.66  &      --        &       --        &   $>$20.05       &       --      &       --        &  2 \\ 
2010-03-05 &  2455260.98  &      --        &       --        &   $>$20.45       &       --      &       --        &  2 \\ 
2010-03-16 &  2455271.81  &      --        &       --        &  20.56 (0.63)    &       --      &       --        &  2 \\ 
\hline                                                                                       
\end{tabular}                                                                               
\end{table*} 
\begin{table*} 
\contcaption{} 
	\begin{tabular}{CCCCCCCc}  
		\hline 
		Date  & JD & $u$ & $g$ & $r$ & $i$ & $z$ & Instrument\\ 
		\hline 
2010-03-25 &  2455280.76  &      --        &       --        &   $>$20.56       &       --      &       --        &  2 \\ 
2010-04-12 &  2455298.88  &      --        &       --        &   $>$20.59       &       --      &       --        &  2 \\ 
2010-05-05 &  2455321.84  &      --        &       --        &   $>$20.40       &       --      &       --        &  2 \\ 
2010-05-14 &  2455330.78  &      --        &       --        &   $>$19.99       &       --      &       --        &  2 \\ 
2010-05-15 &  2455331.77  &      --        &       --        &   $>$20.28       &       --      &       --        &  2 \\ 
2010-05-17 &  2455333.81  &      --        &       --        &   $>$20.71       &       --      &       --        &  2 \\ 
2010-05-25 &  2455341.79  &      --        &       --        &   $>$20.36       &       --      &       --        &  2 \\ 
2010-06-02 &  2455349.18  &      --        &       --        &   $>$21.08       &       --      &       --        &  3 \\ 
2010-06-08 &  2455355.68  &      --        &       --        &   $>$20.06       &       --      &       --        &  2 \\ 
2010-06-12 &  2455359.71  &      --        &       --        &   $>$20.33       &       --      &       --        &  2 \\ 
2011-02-23 &  2455615.97  &      --        &       --        &   $>$20.58       &       --      &       --        &  2 \\ 
2011-03-08 &  2455628.85  &      --        &       --        &   $>$20.30       &       --      &       --        &  2 \\ 
2011-03-14 &  2455634.85  &      --        &       --        &   $>$20.65       &       --      &       --        &  2 \\ 
2011-03-27 &  2455647.77  &      --        &       --        &   $>$20.65       &       --      &       --        &  2 \\ 
2011-04-08 &  2455659.80  &      --        &       --        &   $>$20.85       &       --      &       --        &  2 \\ 
2011-04-28 &  2455679.87  &      --        &       --        &   $>$20.74       &       --      &       --        &  2 \\ 
2011-06-10 &  2455722.74  &      --        &       --        &   $>$20.68       &       --      &       --        &  2 \\ 
2012-01-29 &  2455955.91  &      --        &       --        &  20.60 (0.50)    &       --      &       --        &  2 \\ 
2012-02-24 &  2455981.86  &      --        &       --        &  20.47 (0.42)    &       --      &       --        &  2 \\ 
2012-03-15 &  2456001.96  &      --        &       --        &  20.89 (0.44)    &       --      &       --        &  2 \\ 
2012-03-22 &  2456008.90  &      --        &       --        &  20.20 (0.25)    &       --      &       --        &  2 \\ 
2012-03-29 &  2456015.84  &      --        &       --        &  20.21 (0.40)    &       --      &       --        &  2 \\ 
2012-04-16 &  2456033.86  &      --        &  21.636 (0.173) &  20.911 (0.084)  &       --      &       --        &  4 \\ 
2012-04-20 &  2456037.72  &      --        &       --        &   $>$20.59       &       --      &       --        &  2 \\ 
2012-05-18 &  2456065.77  &      --        &       --        &   $>$20.62       &       --      &       --        &  2 \\ 
2012-06-10 &  2456088.68  &      --        &       --        &   $>$20.48       &       --      &       --        &  2 \\ 
2012-06-18 &  2456096.72  &      --        &       --        &   $>$20.55       &       --      &       --        &  2 \\ 
2013-01-22 &  2456314.86  &      --        &       --        &   $>$20.80       &       --      &       --        &  2 \\ 
2013-03-02 &  2456353.98  &      --        &       --        &   $>$20.37       &       --      &       --        &  2 \\ 
2013-03-17 &  2456368.87  &      --        &       --        &   $>$20.92       &       --      &       --        &  2 \\ 
2013-04-12 &  2456394.84  &      --        &       --        &   $>$18.92       &       --      &       --        &  2 \\ 
2013-04-13 &  2456395.85  &      --        &       --        &   $>$20.24       &       --      &       --        &  2 \\ 
2013-04-21 &  2456403.72  &      --        &       --        &   $>$20.33       &       --      &       --        &  2 \\ 
2013-05-04 &  2456416.70  &      --        &       --        &   $>$20.41       &       --      &       --        &  2 \\ 
2013-06-05 &  2456448.77  &      --        &       --        &   $>$20.66       &       --      &       --        &  2 \\ 
2013-06-18 &  2456461.75  &      --        &       --        &   $>$20.59       &       --      &       --        &  2 \\ 
2014-01-02 &  2456659.96  &      --        &       --        &   $>$20.60       &       --      &       --        &  2 \\ 
2014-02-09 &  2456698.14  &      --        &  $>$21.64       &          --      &       --      &  $>$20.32       &  4 \\ 
2014-02-21 &  2456710.13  &      --        &       --        &   $>$21.68       &       --      &  $>$20.75       &  4 \\ 
2014-03-09 &  2456725.88  &      --        &       --        &   $>$20.67       &       --      &       --        &  2 \\ 
2014-03-18 &  2456735.04  &      --        &       --        &          --      & 21.442 (0.288)&       --        &  4 \\ 
2014-03-26 &  2456742.79  &      --        &       --        &   $>$20.70       &       --      &       --        &  2 \\ 
2014-03-27 &  2456743.91  &      --        &  21.630 (0.174) &          --      &       --      &       --        &  4 \\ 
2014-04-01 &  2456748.75  &      --        &       --        &   $>$20.88       &       --      &       --        &  2 \\ 
2014-04-09 &  2456756.84  &      --        &       --        &   $>$20.62       &       --      &       --        &  2 \\ 
2014-04-25 &  2456772.84  &      --        &       --        &   $>$20.76       &       --      &       --        &  2 \\ 
2014-05-03 &  2456780.81  &      --        &       --        &   $>$20.33       &       --      &       --        &  2 \\ 
2014-05-26 &  2456803.78  &      --        &       --        &   $>$20.77       &       --      &       --        &  2 \\ 
2014-06-06 &  2456814.77  &      --        &       --        &   $>$20.29       &       --      &       --        &  2 \\ 
2014-06-21 &  2456829.77  &      --        &       --        &   $>$20.28       &       --      &       --        &  2 \\ 
2015-01-18 &  2457040.85  &      --        &       --        &   $>$20.57       &       --      &       --        &  2 \\ 
2015-01-25 &  2457047.83  &      --        &       --        &   $>$20.49       &       --      &       --        &  2 \\ 
2015-02-09 &  2457062.94  &      --        &       --        &   $>$20.58       &       --      &       --        &  2 \\ 
2015-02-10 &  2457064.17  &      --        &       --        &          --      &       --      &  $>$19.13       &  4 \\ 
2015-02-11 &  2457065.17  &      --        &       --        &          --      &       --      &  $>$20.22       &  4 \\ 
2015-02-19 &  2457072.96  &      --        &       --        &   $>$20.65       &       --      &       --        &  2 \\ 
2015-03-03 &  2457085.11  &      --        &       --        &          --      & $>$21.05      &       --        &  4 \\ 
2015-03-10 &  2457092.01  &      --        &       --        &   $>$20.57       &       --      &       --        &  2 \\ 
2015-03-17 &  2457098.88  &      --        &       --        &   $>$20.60       &       --      &       --        &  2 \\ 
2015-03-24 &  2457105.78  &      --        &       --        &   $>$20.62       &       --      &       --        &  2 \\ 
2015-04-10 &  2457122.76  &      --        &       --        &   $>$20.65       &       --      &       --        &  2 \\ 
\hline                                                                                       
\end{tabular}                                                                               
\end{table*} 
\begin{table*} 
\contcaption{} 
	\begin{tabular}{CCCCCCCc}  
		\hline 
		Date  & JD & $u$ & $g$ & $r$ & $i$ & $z$ & Instrument\\ 
		\hline 
2015-04-16 &  2457128.89  &      --        &       --        &   $>$20.55       &       --      &       --        &  2 \\ 
2015-04-23 &  2457135.79  &      --        &       --        &  20.70 (0.42)    &       --      &       --        &  2 \\ 
2015-05-19 &  2457161.73  &      --        &       --        &  20.10 (0.40)    &       --      &       --        &  2 \\ 
2015-05-25 &  2457167.66  &      --        &       --        &  20.59 (0.33)    &       --      &       --        &  2 \\ 
2015-06-15 &  2457188.67  &      --        &       --        &   $>$20.90       &       --      &       --        &  2 \\ 
2015-06-29 &  2457202.81  &      --        &       --        &       --         &   $>$21.70    &       --        &  4 \\ 
2015-06-30 &  2457203.79  &      --        &       --        &       --         &   $>$21.53    &       --        &  4 \\ 
2015-12-25 &  2457382.12  &      --        &       --        &       --         &   $>$20.31    &       --        &  4 \\ 
2016-01-18 &  2457406.03  &      --        &       --        &   $>$20.68       &       --      &       --        &  2 \\ 
2016-01-30 &  2457418.01  &      --        &       --        &   $>$20.63       &       --      &       --        &  2 \\ 
2016-01-31 &  2457419.05  &      --        &       --        &  21.672 (0.243)  &       --      &       --        &  4 \\ 
2016-02-06 &  2457424.99  &      --        &       --        &   $>$20.39       &       --      &       --        &  2 \\ 
2016-02-13 &  2457432.00  &      --        &       --        &  20.50 (0.35)    &       --      &       --        &  2 \\ 
2016-02-28 &  2457446.97  &      --        &       --        &  20.77 (0.72)    &       --      &       --        &  2 \\ 
2016-02-28 &  2457447.13  &      --        &       --        &  20.729 (0.170)  &       --      &       --        &  4 \\ 
2016-03-04 &  2457452.14  &      --        &       --        &       --         &       --      & 20.307 (0.245)  &  4 \\ 
2016-03-12 &  2457459.95  &      --        &       --        &  20.83 (0.55)    &       --      &       --        &  2 \\ 
2016-03-18 &  2457465.93  &      --        &       --        &  20.621 (0.132)  &       --      &       --        &  4 \\ 
2016-03-18 &  2457465.94  &      --        &       --        &  20.64 (0.41)    &       --      &       --        &  2 \\ 
2016-03-18 &  2457465.94  &      --        &       --        &  20.656 (0.139)  &       --      &       --        &  4 \\   
2016-03-28 &  2457475.90  &      --        &       --        &  19.945 (0.053)  &       --      &       --        &  4 \\   
2016-03-31 &  2457478.77  &      --        &       --        &  19.78 (0.40)    &       --      &       --        &  2 \\ 
2016-04-06 &  2457484.76  &      --        &       --        &  19.85 (0.33)    &       --      &       --        &  2 \\ 
2016-04-27 &  2457505.72  &      --        &       --        &  19.14 (0.21)    &       --      &       --        &  2 \\ 
2016-05-05 &  2457513.90  &      --        &  $>$19.43       &    --           &       --       &       --        &  5 \\   
2016-05-08 &  2457516.84  &      --        &       --        &  19.44 (0.25)    &       --      &       --        &  2 \\ 
2016-05-20 &  2457528.86  &      --        &       --        &  $>$18.75        &       --      &       --        &  5 \\ 
2016-05-24 &  2457532.87  &      --        &       --        &  16.909 (0.071)  &       --      &       --        &  5 \\ 
2016-05-24 &  2457532.91  &      --        &       --        &       --        &  17.167 (0.022)&       --        &  4 \\    
2016-05-28 &  2457536.86  &      --        &       --        &  16.438 (0.080) &       --       &       --        &  5 \\ 
2016-06-02 &  2457541.80  &      --        &       --        &  16.385 (0.197) &       --       &       --        &  2 \\ 
2016-06-02 &  2457541.80  &      --        &       --        &  16.434 (0.030) &  16.498 (0.035)&       --        &  6 \\ 
2016-06-02 &  2457542.48  &      --        &       --        &  16.355 (0.165) &       --       &       --        &  7 \\ 
2016-06-03 &  2457542.85  &      --        &  16.412 (0.049) &    --           &       --       &       --        &  5 \\ 
2016-06-07 &  2457546.85  &      --        &  16.569 (0.043) &    --           &       --       &       --        &  5 \\    
2016-06-12 &  2457552.42  & 17.048 (0.015) &  16.838 (0.014) &  16.710 (0.017) &  16.836 (0.017)&       --        &  8 \\ 
2016-06-13 &  2457552.81  &      --        &  16.820 (0.026) &  16.756 (0.061) &  16.814 (0.050)&       --        &  9 \\ 
2016-06-14 &  2457553.56  & 17.123 (0.033) &  16.787 (0.020) &  16.751 (0.017) &  16.830 (0.021)&  16.960 (0.014) &  7 \\ 
2016-06-14 &  2457553.80  &      --        &  16.826 (0.037) &  16.779 (0.074) &  16.853 (0.064)&       --        &  9 \\ 
2016-06-14 &  2457554.49  & 17.180 (0.025) &  16.847 (0.024) &  16.769 (0.019) &  16.888 (0.015)&       --        &  8 \\ 
2016-06-16 &  2457555.83  &      --        &  16.936 (0.073) &  16.867 (0.079) &  16.912 (0.052)&       --        &  9 \\ 
2016-06-17 &  2457557.41  & 17.553 (0.067) &  16.954 (0.036) &  16.957 (0.046) &  17.036 (0.037)&  17.251 (0.036) &  10 \\ 
2016-06-18 &  2457557.81  &      --        &       --        &  $>$17.06       &       --       &       --        &  5 \\ 
2016-06-18 &  2457558.48  & 17.609 (0.019) &  17.037 (0.022) &  17.003 (0.026) &  17.066 (0.025)&       --        &  8 \\ 
2016-06-21 &  2457561.44  & 18.172 (0.030) &  17.382 (0.014) &  17.171 (0.020) &  17.296 (0.031)&       --        &  8 \\ 
2016-06-22 &  2457561.82  &      --        &       --        &  17.200 (0.039) &       --       &       --        &  5 \\ 
2016-06-22 &  2457562.43  & 18.228 (0.050) &  17.431 (0.065) &  17.234 (0.036) &  17.368 (0.031)&  17.426 (0.028) &  10 \\ 
2016-06-26 &  2457565.82  &      --        &       --        &  17.451 (0.045) &       --       &       --        &  5 \\ 
2016-07-01 &  2457571.50  &      --        &       --        &  17.688 (0.193) &       --       &       --        &  7 \\ 
2016-07-03 &  2457572.69  &      --        &       --        &  17.825 (0.187) &       --       &       --        &  6 \\ 
2016-07-04 &  2457573.51  & 19.623 (0.033) &  18.372 (0.033) &  17.877 (0.024) &  17.948 (0.021)&  17.912 (0.020) &  7 \\ 
2016-07-04 &  2457573.67  &      --        &       --        &  17.912 (0.197) &       --       &       --        &  6 \\ 
2016-07-13 &  2457582.78  &      --        &  18.481 (0.152) &    --           &       --       &       --        &  5 \\ 
2016-07-14 &  2457583.68  &      --        &       --        &  18.078 (0.201) &       --       &       --        &  6 \\ 
2016-07-15 &  2457585.49  & 20.259 (0.118) &  18.674 (0.038) &  18.105 (0.021) &  18.047 (0.018)&  18.074 (0.036) &  7 \\ 
2016-07-17 &  2457587.41  &      --        &       --        &  18.102 (0.084) &       --       &       --        & 11 \\ 
2016-07-21 &  2457591.41  &      --        &  18.793 (0.027) &  18.203 (0.022) &  18.153 (0.031)&  18.087 (0.035) &  7 \\ 
2016-07-28 &  2457598.43  &      --        &  18.991 (0.028) &  18.291 (0.014) &  18.275 (0.014)&  18.177 (0.029) &  7 \\ 
2016-07-29 &  2457599.43  &      --        &       --        &  18.317 (0.349) &       --       &       --        &  7$^\ddag$ \\ 
2016-07-30 &  2457599.77  &      --        &       --        &  18.333 (0.198) &       --       &       --        &  5 \\ 
2016-08-04 &  2457605.43  & 21.398 (0.127) &  19.600 (0.020) &  18.470 (0.019) &  18.399 (0.018)&  18.446 (0.037) &  7 \\ 
\hline                                                                                       
\end{tabular}                                                                               
\end{table*} 
\begin{table*} 
\contcaption{} 
	\begin{tabular}{CCCCCCCc}  
		\hline 
		Date  & JD & $u$ & $g$ & $r$ & $i$ & $z$ & Instrument\\ 
		\hline 
2016-08-06 &  2457606.79  &      --        &   $>$19.29      &    --           &       --       &       --        &  5 \\  
2016-08-10 &  2457611.42  &      --        &  20.233 (0.058) &  19.176 (0.024) &  19.202 (0.026)&  19.106 (0.041) &  7 \\ 
2016-08-18 &  2457619.40  &      --        &  20.601 (0.055) &  19.529 (0.024) &  19.578 (0.029)&  19.488 (0.036) &  7 \\ 
2016-08-19 &  2457620.37  &      --        &       --        &  19.599 (0.085) &       --       &       --        & 11 \\ 
2016-08-22 &  2457623.34  &      --        &       --        &  19.650 (0.040) &  19.771 (0.062)&  19.532 (0.079) & 10 \\  
2016-08-23 &  2457624.31  &  $>$20.91      &  20.682 (0.244) &       --        &       --       &       --        & 10 \\ 
2016-08-27 &  2457628.31  &      --        &       --        &  19.778 (0.065) &  19.870 (0.088)&  19.652 (0.064) & 10 \\ 
2016-08-27 &  2457628.39  &      --        &  20.772 (0.121) &       --        &       --       &       --        &  7 \\   
2016-12-23 &  2457746.16  &      --        &  $>$20.91       &       --        &       --       &       --        &  5 \\ 
2016-12-29 &  2457751.74  &      --        &  22.693 (0.081) &  20.898 (0.040) &  21.872 (0.072)&  21.639 (0.144) &  7 \\ 
2016-12-31 &  2457754.16  &      --        &  $>$20.76       &       --        &       --       &       --        &  5 \\ 
2017-01-05 &  2457758.77  &  $>$22.36      &       --        &       --        &       --       &       --        &  7 \\ 
2017-01-20 &  2457773.61  &      --        &       --        &  20.947 (0.299) &       --       &       --        & 11 \\ 
2017-01-24 &  2457777.53  &      --        &       --        &       --        &  22.106 (0.275)&       --        & 10 \\ 
2017-01-25 &  2457778.65  &      --        &  22.880 (0.296) &  20.969 (0.128) &  22.130 (0.324)&  21.881 (0.460) & 10 \\ 
2017-02-19 &  2457803.74  &      --        &  23.254 (0.263) &  21.245 (0.032) &  22.341 (0.082)&   $>$21.92      &  7 \\ 
2017-03-26 &  2457838.63  &      --        &  23.335 (0.097) &  21.437 (0.042) &  22.534 (0.086)&  22.208 (0.182) &  7 \\ 
        	\hline 
        \end{tabular} 
\begin{flushleft} 
1 = 2.54~m INT + WFC; 2 = CRTS Telescope; 3 = PTF Telescope ($R$ filter, Sloan $r$ calibrated); 4 = PS1; 5 = ATLAS Telescopes (``open'' filter, Sloan $r$ calibrated);  6 = Meade 10'' LX-200 Telescope + Apogee Alte F-47 CCD;  
7 = 2.56~m NOT + ALFOSC; 8 = 2.0~m LT + IO:O; 9 = 0.51~m Iowa Robotic Telescope + Apogee Alta F-47 CCD; 10 = 1.82~m Copernico Telescope + AFOSC; 11 = 10.4~m GTC + OSIRIS.\\  
 
$^\ddag$ = unfiltered observation, Sloan $r$  calibrated; $^\oplus$ = $Y$ filter, Sloan $r$  calibrated; $^\dag$ = $R$ filter, Sloan $r$  calibrated.  
\end{flushleft} 
\end{table*}

\begin{table*} 
	\centering 
	\caption{Near-infrared photometry of SN~2016bdu. } 
	\label{tab:2016bdu_NIR} 
	\begin{tabular}{CCCCC}  
		\hline 
		Date & JD & $J$ & $H$ & $K$ \\ 
		\hline 
2016-06-11 & 2457551.48 &  16.41 (0.18)  & 16.32 (0.25)  & 16.05 (0.23) \\ 
2016-06-23 & 2457562.50 &  16.73 (0.19)  & 16.63 (0.24)  & 16.35 (0.23) \\ 
2016-07-14 & 2457584.44 &  17.18 (0.21)  & 17.12 (0.21)  & 16.82 (0.16) \\ 
2016-07-27 & 2457597.40 &  17.34 (0.20)  & 17.21 (0.19)  & 16.94 (0.16) \\ 
2016-08-11 & 2457612.43 &  18.39 (0.16)  & 18.15 (0.30)  & 17.64 (0.25) \\ 
2016-08-22 & 2457623.38 &  18.80 (0.22)  & 18.51 (0.39)  & 17.92 (0.30) \\ 
2017-01-10 & 2457763.68 &  20.60 (0.24)  & 20.67 (0.38)  & 20.31 (0.24) \\ 
2017-02-07 & 2457791.63 &  21.21 (0.32)  & 20.95 (0.46)  & 20.42 (0.39) \\ 
2017-03-24 & 2457837.45 &  21.76 (0.25)  & 21.18 (0.45)  & 20.86 (0.47) \\ 
        	\hline 
        \end{tabular} 
\begin{flushleft} 
Observations obtained using the 2.56~m NOT + NOTCam. 
\end{flushleft} 
\end{table*}

\section{Photometric data for SN~2005gl} \label{app1b} 
 
We remeasured the unfiltered KAIT images of SN~2005gl previously published by \citet{gal07}, and collected new data from amateur astronomers. 
 These unfiltered photometric data were calibrated to the Johnson-Cousins $R$ band. The $R$-band calibration, in fact, provided more accurate results because the quantum  
efficiency curves of the CCDs used in these observations typically peaked at $\lambda \geq 6000$~\AA. 
For the calibration, we used the $R$-band magnitudes of comparison stars from \citet{gal07}. 
There is a significant offset from the magnitudes provided by \citet{gal07}. This disagreement is likely due to the fact that 
we did not use Stars 4 and 7 of \citet{gal07} for our calibration. These two stars were rejected because the $V-R$ colours reported 
in Table 2 of \citet{gal07} are extremely blue ($-0.26$ mag and $-0.22$ mag for Stars 4 and 7, respectively), in contrast with 
the relatively red colours reported for these two stars in the Sloan catalogue ($g-r \approx 1.2$ mag). 
The photometry for both datasets was obtained with the template-subtraction method.  
 
\begin{table*} 
	\centering 
	\caption{Photometry of SN~2005gl (unfiltered KAIT and amateur), rescaled to Johnson-Cousins $R$ (Vega mag).} 
	\label{tab:2005gl} 
	\begin{tabular}{cCCc}  
		\hline 
		Date & Average JD & $R$ & Instrument \\ 
		\hline 
stack 1998-10-20 to 1999-01-12 & 2451149.75 & $>$20.90 & 1 \\ 
stack 1999-06-29 to 2000-01-06 & 2451460.42 & $>$20.84 & 1 \\ 
stack 2000-07-10 to 2000-01-13 & 2451819.98 & $>$21.11 & 1 \\ 
stack 2001-07-31 to 2001-12-15 & 2452179.42 & $>$20.87 & 1 \\ 
stack 2002-07-18 to 2003-01-08 & 2452552.11 & $>$21.13 & 1 \\ 
stack 2003-07-15 to 2003-02-22 & 2452898.56 & $>$21.37 & 1 \\ 
2005-07-22  & 2453573.94   &   19.99  (0.45) & 1 \\ 
2005-07-28  & 2453579.97   &   $>$19.73      & 1 \\ 
2005-08-03  & 2453585.94   &   $>$20.22      & 1 \\ 
2005-08-09  & 2453591.95   &   $>$20.20      & 1 \\ 
2005-08-15  & 2453597.86   &   $>$17.39      & 1 \\ 
2005-08-20  & 2453602.89   &   $>$19.89      & 1 \\ 
2005-08-26  & 2453608.87   &   $>$20.42      & 1 \\ 
stack 2005-08-03 to 2005-09-14 & 2453610.36 & $>$20.84 & 1 \\ 
2005-08-31  & 2453613.85   &   $>$20.50      & 1 \\ 
2005-09-09  & 2453622.81   &   $>$20.62      & 1 \\ 
2005-09-14  & 2453627.80   &   $>$20.20      & 1 \\ 
2005-10-02  & 2453645.80   &   $>$16.97      & 1 \\ 
2005-10-05  & 2453648.60   &   18.81 (0.66)  & 2 \\ 
2005-10-06  & 2453649.60   &   18.22 (0.39)  & 2 \\ 
2005-10-08  & 2453651.81   &   $>$17.53      & 1 \\ 
2005-10-12  & 2453656.09   &   17.61 (0.15)  &   2 \\  
2005-10-13  & 2453656.73   &   17.50 (0.11)  &   1 \\ 
2005-10-13  & 2453657.05   &   17.46 (0.12)  &   2 \\  
2005-10-15  & 2453659.37   &   17.19 (0.12)  &   3 \\ 
2005-10-16  & 2453660.05   &   17.09 (0.11)  &   2 \\  
2005-10-18  & 2453661.92   &   16.91 (0.09)  &   2 \\  
2005-10-20  & 2453663.82   &   16.75 (0.13)  &   1 \\ 
2005-10-20  & 2453664.06   &   16.76 (0.09)  &   2 \\  
2005-10-25  & 2453668.74   &   16.80  (0.13) &   1 \\ 
2005-10-25  & 2453669.39   &   16.81  (0.04) &   4 \\ 
2005-10-27  & 2453670.89   &   16.82  (0.10) &   2 \\  
2005-10-27  & 2453671.06   &   16.82  (0.06) &   2 \\  
2005-10-27  & 2453671.29   &   16.84  (0.09) &   5 \\ 
2005-10-27  & 2453671.31   &   16.85  (0.06) &   4 \\ 
2005-10-31  & 2453674.79   &   16.94  (0.07) &   1 \\ 
2005-11-06  & 2453680.24   &   17.04  (0.12) &   1 \\ 
2005-11-07  & 2453682.35   &   17.06  (0.06) &   4 \\ 
2005-11-07  & 2453682.37   &   17.14  (0.10) &   5 \\ 
2005-11-15  & 2453689.92   &   17.24  (0.14) &   2 \\ 
2005-11-17  & 2453692.34   &   17.26  (0.09) &   4 \\ 
2005-11-20  & 2453694.75   &   17.36  (0.14) &   1 \\ 
2005-11-20  & 2453695.26   &   17.37  (0.06) &   4 \\ 
2005-11-20  & 2453695.37   &   17.36  (0.05) &   3 \\  
2005-11-26  & 2453700.80   &   17.57  (0.21) &   1 \\ 
2005-11-30  & 2453705.33   &   17.72  (0.19) &   5 \\ 
2005-12-07  & 2453711.68   &   18.24  (0.11) &   1 \\ 
2005-12-18  & 2453723.24   &   18.83  (0.26) &   4 \\  
2005-12-21  & 2453726.23   &   19.18  (0.21) &   5 \\ 
2005-12-25  & 2453729.67   &   19.87  (0.35) &   1 \\ 
2005-12-26  & 2453731.27   &   19.94  (0.33) &   4 \\ 
2006-01-06  & 2453742.26   &   20.20  (0.45) &   3 \\ 
2006-01-16  & 2453751.61   &   20.58  (0.53) &   1 \\ 
2006-07-08  & 2453924.91   &   $>$20.40      &   1 \\ 
2006-07-15  & 2453931.98   &   $>$20.35      &   1 \\ 
       	\hline 
        \end{tabular} 
\begin{flushleft} 
1 = KAIT data; 2 = 0.28~m C11 reflector + SBIG ST-9E CCD camera (Obs. Y. Sano, Nayoro, Japan); 3 = 0.356~m Meade LX200 Telescope + SBIG ST-9XE CCD camera (Obs. E. Prosperi, Larciano, Italy); 4 = 0.28~m C11 reflector + SBIG ST-8XME Kaf1602E CCD camera (Obs. J. Nicolas; Vallauris, France); 5 = 0.28~m C11 reflector + SBIG ST-8XME Kaf1602E CCD camera (Obs. J.-M. Llapasset; Perpignan, France). 
\end{flushleft} 
\end{table*}

\bsp	
\label{lastpage} 
\end{document}